\documentclass[twocolumn,trackchanges]{aastex631}
\usepackage{amsmath}
\newcommand{\diff}[1]{\rm{d} #1}

\usepackage{subfigure}
\usepackage{soul}
\usepackage{cancel}

\begin{document}

\title{Exploring the Future of Soft X-ray Polarimetry: the Capabilities of the REDSoX Instrument for 
XDINS and Magnetar Studies}

\author[0000-0002-5004-3573]{Ruth M. E. Kelly}
\affiliation{Mullard Space Science Laboratory, University College London \\ Holmbury St Mary, Dorking, Surrey RH5 6NT, UK}
\affiliation{MIT Kavli Institute for Astrophysics and Space Research, Massachusetts Institute of Technology \\ 77 Massachusetts Avenue, Cambridge, MA 02139, USA}

\author{Herman L. Marshall}
\affiliation{MIT Kavli Institute for Astrophysics and Space Research, Massachusetts Institute of Technology \\ 77 Massachusetts Avenue, Cambridge, MA 02139, USA}


\author{Silvia Zane}
\affiliation{Mullard Space Science Laboratory, University College London \\ Holmbury St Mary, Dorking, Surrey RH5 6NT, UK}

\author{Nabil Brice}
\affiliation{Centre for Astrophysics Research,
University of Hertfordshire \\ College Lane, Hatfield, AL10 9AB, UK}

\author{Swati Ravi}
\affiliation{MIT Kavli Institute for Astrophysics and Space Research, Massachusetts Institute of Technology \\ 77 Massachusetts Avenue, Cambridge, MA 02139, USA}

\author{Roberto Turolla}
\affiliation{Universit\`a di Padova, Dipartimento di Fisica e Astronomia \\ via Marzolo 8, I-35131 Padova, Italy}
\affiliation{Mullard Space Science Laboratory, University College London \\ Holmbury St Mary, Dorking, Surrey RH5 6NT, UK}

\author{Roberto Taverna}
\affiliation{Universit\`a di Padova, Dipartimento di Fisica e Astronomia \\ via Marzolo 8, I-35131 Padova, Italy}

\author{Denis Gonz\'alez-Caniulef}
\affiliation{Institut de Recherche en Astrophysique et Plan\'etologie \\ 9 avenue du Colonel Roche, BP 44346 F-31028, Toulouse CEDEX 4, France}



\begin{abstract}
X-ray polarimetry offers a unique window into neutron star physics and can provide answers to questions that cannot otherwise be probed.
The up-and-coming REDSoX sounding rocket mission will be the first experiment equipped with a detector able to explore polarized X-rays below $1$ keV, observing in the $0.2$--$0.4$ keV range.
Although REDSoX will only be capable of short, one-off observations, it will crucially test the instrument performance. In this paper we investigate how a fully-fledged orbital mission with longer lifetime, based on an instrument design similar to REDSoX, will allow us to study thermal emission from the X-ray dim isolated neutron stars (XDINSs) and magnetars, probing their magnetic field and the physics of their outer surface layers,
including vacuum effects and QED mode conversion at the vacuum resonance. We discuss the potentially observable features for promising values of the star's surface temperature, magnetic field, and viewing geometry. Assuming emission from the whole surface, we find that, for a source with a magnetic field $B =5\times10^{13}\, \mathrm{G}$ and surface temperature $T \approx 10^7 \, \mathrm{K}$, the instrument can resolve a proton-cyclotron absorption feature in the spectrum with high significance when collecting $\approx25,000$ counts across a single observation. Similarly, for a source with $B = 10^{14}\, \mathrm{G}$ and $T \approx 10^7 \, \mathrm{K}$, a switch in the dominant polarization mode,  caused by mode conversion at the vacuum resonance, can be detected by collecting $\approx 25,000$ counts, allowing for a long-sought observational test of the presence of QED effects. We then
present two case studies for XDINS targets: RX J1856.5$-$3754 and RX J0720.4$-$3125. 

\end{abstract}


\section{Introduction} \label{sec:intro}

Polarimetry is a relatively young field in X-ray astronomy but is crucial to our understanding of celestial sources.
Although unquestionably a powerful and useful observational technique, historically X-ray polarimetry has been underutilized in astronomy, mainly due to the lack of technology. The situation has changed very recently: the NASA-ASI Imaging X-ray Polarimetry Explorer (IXPE) space observatory was launched in December 2021 \cite[]{weisskopf_imaging_2022} and it was followed in early 2024 by the Indian Space Research Organisation's X-ray Polarimeter Satellite (XPoSat, \citealt{isro_xposat_2024}). 
IXPE is a mission dedicated to systematically exploring polarized X-rays from the sky in the $2$--$8\,\mathrm{keV}$ energy range, and has observed over 150 targets to date. It
has been the first space mission of this kind and
is capable of performing spatial-, energy-, and time-resolved polarimetry. IXPE flew over 40 years after the first pioneering attempts made in 1971, 1974 and 1975 with the launches of Aerobee-350 \citep{novick_detection_1972}, Ariel-5 \citep{smith_ariel_1976} and the 8th Orbiting Solar Observatory \cite[OSO-8,][]{weisskopf_x-ray_1976}.
Following the ongoing success of IXPE, and the scientific contributions of the PoGO+ \citep{chauvin_calibration_2017} and PolarLight \citep{feng_polarlight_2019} instruments, there has been an increase in interest in spaceborne polarimetric instruments. In January 2024, XPoSat was launched with the Polarimeter Instrument in X-rays (POLIX) onboard, capable of observing X-ray polarization in the $8$--$30\,\mathrm{keV}$ range \citep{isro_xposat_2024}. The Chinese Academy of Sciences
is expected to launch the enhanced X-ray Timing and Polarimetry (eXTP) mission by 2030, with an onboard polarimeter operating in the same energy band as IXPE, $2$--$8 \ \mathrm{keV}$, but with much higher sensitivity \cite[]{zhang_enhanced_2019}. 

However, to fully understand the physical origin of the X-ray polarization from astrophysical sources, observations covering a broader range of energies are necessary. In particular, the soft X-ray range, below that covered by IXPE, still remains unexplored, despite potentially being key to understand fascinating phenomena ranging from vacuum birefringence \citep{heisenberg_folgerungen_1936} to the magnetic field dependence of blazar jets \citep{blandford_relativistic_2019}.

A first attempt in this direction will be made with the Rocket Experiment Demonstration of a Soft X-ray Polarimeter (REDSoX), which is scheduled for launch in 2027 and will become the first instrument to observe below $1\,\mathrm{keV}$ \citep{marshall_design_2018}. The rocket experiment will take polarimetric measurements between $0.2$--$0.4 \, \mathrm{keV}$ for $\sim 5$ minutes, with the main goal of testing the capabilities and performances of the instrument. The success of this experiment will open the door to the development of the Globe Orbiting Soft X-ray Polarimeter \cite[GOSoX,][]{marshall_globe_2021}, a fully functioning orbital mission equipped with similar instrument design.  

The potential of an instrument capable of observing polarization in the soft X-ray range is particularly high for X-ray sources such as X-ray Dim Isolated Neutron Stars (XDINSs) and magnetars.
With inferred dipolar magnetic fields in excess of $\sim 10^{13}\ \mathrm{G}$, they are ideal candidates for studying the radiation from the neutron star surface, potentially testing quantum electrodynamics (QED) effects such as vacuum birefringence \citep{heisenberg_folgerungen_1936}.

Magnetars are believed to host complex magnetic fields which are the strongest among all astrophysical objects \cite[see][for a review]{turolla_magnetars_2015}. They are relatively young isolated neutron stars and have recently been observed to produce highly polarized X-ray emission. To date, IXPE has observed five magnetars and has discovered some fascinating polarization properties.
Not only did the highest detected polarization  come from one magnetar source \cite[$\approx 80\%$,][]{zane_strong_2023}, a clear change in polarization angle of  $90^\circ$, showing two different dominant polarization modes across the energy range, was also observed in another source \citep{taverna_polarized_2022}. The observations across all five sources show indications of both atmospheric and condensate components at the surface, as well as reprocessing of emission by resonant Compton scattering \cite[RCS,][]{taverna_polarized_2022, zane_strong_2023, turolla_ixpe_2023, heyl_detection_2024, rigoselli_ixpe_2024, stewart_x-ray_2024}. In one source, evidence of a surface magnetic loop was also discovered \citep{pizzocaro_detailed_2019, heyl_detection_2024}. Models of particle bombardment \cite[][see also \citealt{gonzalez-caniulef_atmosphere_2019}]{zane_strong_2023, kelly_x-ray_2024} and partial mode conversion \citep{lai_ixpe_2023} have also been suggested to explain some of the observed polarization characteristics, although the latter scenario has been challenged by \cite{kelly_x-ray_2024-1}.

XDINSs, also known as the ``Magnificent Seven'', are a class of seven neutron stars, which are among the closest to Earth. These neutron stars emit soft thermal radiation ($kT_\mathrm{BB} \sim 40$--$100 \, \mathrm{eV}$), with little, or no, evidence of a high-energy tail\footnote{A high-energy excess has been recently reported in RX J1856.5$-$3754 \citep{degrandis_two_2022}.}, and are therefore ideal candidates for soft X-ray polarimetry \citep{turolla_isolated_2009, van_kerkwijk_isolated_2007}. The surface temperature distribution is expected to be nonuniform, pointing to complex magnetic field structures \citep{vigano_spectral_2014, popov_probing_2017, de_grandis_x-ray_2021}. Absorption features have been found in six of the seven XDINS sources. Their origin is not yet clear, but they are probably due to atomic transitions or cyclotron absorption. The emission from RX J1856.5$-$3754, the brightest of the XDINSs and the only one without an observed absorption feature \citep{sartore_spectral_2012}, was found to be polarized in the optical band (with a polarization degree, $\mathrm{PD}$, of $16.43 \pm 5.26 \%$), which could be an indication that vacuum birefringence is at work around the star \citep{mignani_evidence_2017}. Although intriguing, it has been noticed that this measurement is significant at the $\sim 3.1\sigma$ level, and further dedicated observations of neutron star surfaces in the X-ray band hold the key to potentially providing a definite test. 

The aim of this work is to explore the potential and capabilities of the REDSoX polarimeter, in view of its potential application within a space mission. 
The paper is laid out as follows. In Section \ref{sec:0.1-1keV} we summarize the polarization signatures expected from atmospheric emission of magnetars and XDINSs in the energy range below $1\ \mathrm{keV}$. 
In Section \ref{sec:application}, we discuss the REDSoX experiment and present some predictions for potential observations of these sources that can be carried out by using the polarimeter in the contest of an orbital space observatory. 
As case studies, we then investigate two XDINSs, RX J1856.5$-$3754 and RX J0720.4$-$3125 (RX~J1856 and RX~J0720 in the following), exploring different parameter combinations, and present our findings in Sections \ref{sec:1856} and \ref{sec:0720}, respectively.
Finally, a discussion and conclusions follow in Section \ref{sec:discussion}.

\section{Surface emission from magnetized neutron stars below 1 keV} \label{sec:0.1-1keV}

Thermal emission from a strongly magnetized isolated neutron star is expected to be due to photons coming directly from its condensed surface or reprocessed by an atmosphere covering the surface \citep{gonzalez_caniulef_polarized_2016, taverna_x-ray_2020}.
As both of these scenarios produce blackbody-like spectra, it can be difficult to disentangle one from the other without an additional observable, such as X-ray polarization.

In the presence of a strong magnetic field, radiation is expected to propagate in two orthogonal linear polarization modes, with the electric vector either parallel or perpendicular to the plane of the local magnetic field and the direction of propagation \citep{gnedin_transfer_1974}. These normal modes, called ordinary (O) and extraordinary (X), are expected to have very different opacities, with the X mode cross sections significantly reduced compared to those of the O mode photons below the electron-cyclotron energy.

The physical state of the outermost layers of a highly magnetized neutron star is still poorly known, and this uncertainty impacts the polarization properties of the thermal radiation we observe. For a long time, it was assumed that neutron stars were covered by a geometrically thin plasma atmosphere, which is expected to emit highly polarized radiation (up to $80\%$ in the X mode),
as a result of the reduced X mode opacity. However, a superstrong magnetic field can dramatically affect the properties of matter. Atoms in such fields can elongate along the field lines, forming molecular chains, and this may result in a phase transition, leaving a condensate on top of a highly magnetized neutron star crust: in place of being covered by a gaseous atmosphere, the neutron star ``condensed surface'' would remain exposed \citep{brinkmann_thermal_1980, lai_hydrogen_1997}. 
Condensed surfaces can form in neutron stars when the temperature is sufficiently low and the magnetic field is strong enough, depending on the chemical composition \cite[see e.g. Figure 1 in ][]{taverna_x-ray_2020}. The emission properties of a condensed surface depend crucially on the assumed form of the matter (plus vacuum) dielectric tensor. Since this is still not fully understood, two limiting cases are usually considered, in which the lattice ions do not move in response to an incident electromagnetic wave or are free to respond as if the lattice is not present: the real optical properties of the medium should be intermediate between the two limits, known as the ``fixed'' and ``free'' ion models, respectively \citep{van_adelsberg_radiation_2005}.

The emission from a condensed surface is expected to have a modest polarization ($\lesssim20 \%$ in the $0.2$--$0.4 \, \mathrm{keV}$), and can be dominated by photons in the X or O mode \citep{gonzalez_caniulef_polarized_2016, taverna_x-ray_2020}.
Then, it follows that the polarization measurement from an isolated neutron star can inform us about the physical state of the external layers of the source, indicating if the emission has a condensate or atmospheric origin.

In this and in the following section, we focus on models based on atmospheric emission. We will then add the results corresponding to a scenario of emission from a condensed surface when we deal with the two case studies (see \S~\ref{sec:casestudies}). Our spectral atmospheric models are obtained by re-adapting the numerical code by \cite{lloyd_model_2003}. The radiation transport for the two normal modes is solved in a plane-parallel, geometrically thin slab of a fully ionized
hydrogen atmosphere, along with the energy balance equation, assuming hydrostatic, radiative, and local thermal equilibrium conditions. The radiative transfer equations are solved on optical depth, frequency, and photon direction meshes, 
and each atmospheric slab is characterized by four parameters: the magnetic field strength and inclination with respect to the slab normal, the effective temperature, and the surface gravity \cite[see][also \citealt{kelly_x-ray_2024, kelly_x-ray_2024-1} for details]{lloyd_model_2003}. In this section, we present some examples computed by assuming a star mass $M = 1.4\,M_{\odot}$,  a radius $R=12 \, \mathrm{km}$, and an effective temperature of $T_\mathrm{eff} = 10^7 \, \mathrm{K}$. 

\subsection{Emission from a single surface patch}
\label{singlepatch}

Under our assumption of a fully ionized H atmosphere, one prominent feature expected in the spectra of highly magnetized neutron stars is the absorption feature at the proton cyclotron resonance energy,  $E_\mathrm{c,p}\simeq 0.63 y_\mathrm{G} (B/10^{14}\mathrm{G}) \ \mathrm{keV}$, where $y_\mathrm{G} = \sqrt{1-(2GM/c^2R)} \approx \sqrt{1-2.98(M/R)}$ is the gravitational redshift factor for a neutron star with mass $M$ and radius $R$ \citep{zane_proton_2001}. 
For magnetic field strengths of the order of $10^{13}$--$10^{14} \ \mathrm{G}$, in the range of the inferred dipole field strengths of XDINSs and some magnetars,  we then expect such proton cyclotron lines to appear in the spectrum  at energies below $1 \ \mathrm{keV}$.  

\begin{figure}
    \centering
    \includegraphics[width=1\linewidth]{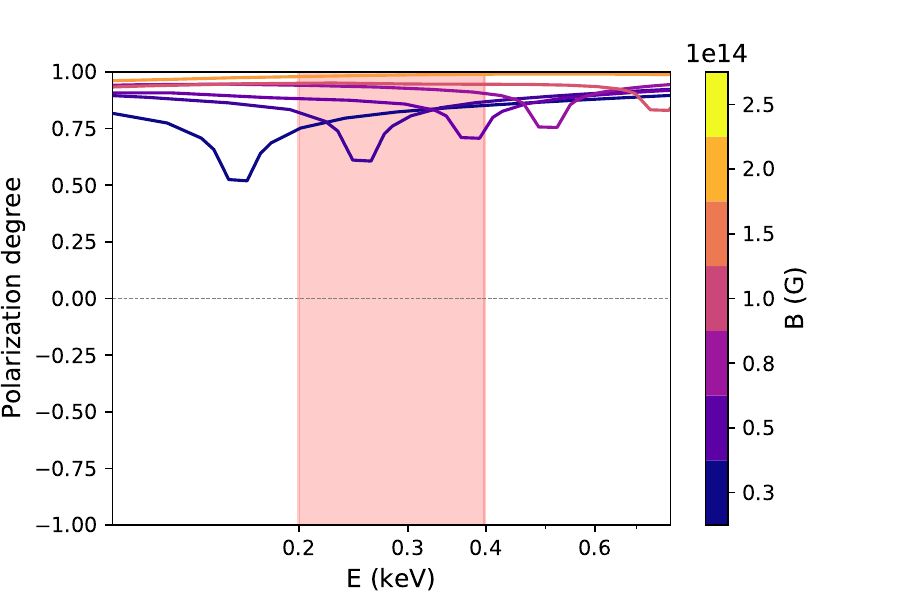}
    \caption{Polarization degree as a function of energy for a pure plasma atmosphere with an orthogonal magnetic field for different values of the field strength in the range $3\times10^{13}$--$2.5\times10^{14} \ \mathrm{G}$. Positive (negative) polarization degrees correspond to a X (O) mode dominated emission. The shaded region highlights the $0.2$--$0.4\,\mathrm{keV}$ range in which REDSoX and GOSoX are expected to observe. The horizontal axis shows the energy as  measured at infinity, i.e. redshifted by the factor $y_\mathrm{G}$ with respect to the photon energy at the neutron star surface. Vacuum effects in the atmosphere are neglected.}
    \label{fig:standard}
\end{figure}

Figure \ref{fig:standard} shows the degree of polarization as a function of energy, at the star surface, but redshifted to be the corresponding energy as measured by a distant observer, for some values of the magnetic field strength in this range and for the case of a pure plasma atmosphere. Here 
$\mathrm{PD} = ({J^\mathrm X_\nu-J^\mathrm O_\nu})/({J^\mathrm X_\nu+J^\mathrm O_\nu})$, where $J^\mathrm{X}_\nu$ ($J^\mathrm{O}_\nu$) is the mean monochromatic intensity of X (O) mode photons, therefore a positive (negative) degree of polarization indicates emission dominated by X (O) photons. The energy on the x-axis has been redshifted (by the gravitational redshift factor $y_\mathrm{G}$) to allow a more immediate comparison of the results with the $0.2$--$0.4 \ \mathrm{keV}$ energy range of the REDSoX instrument. The proton cyclotron line is present in the polarization spectrum, and it falls 
in the energy range of the detector
for magnetic field strengths $3\times10^{13}\,\mathrm{G} \lesssim  B \lesssim 10^{14} \ \mathrm{G}$.
Clearly, the detection of such a polarization dip from a source would confirm the nature of the absorption feature in the spectrum and allow us to constrain the magnetic field strength.

However, QED effects on radiation transfer can induce different energy dependencies and features in the polarization signal, which can be observed in the REDSoX band.
The presence of magnetic fields with strength around (or in excess of) the quantum critical field ($B_\mathrm{Q}\simeq4.4\times10^{13}\,\mathrm{G}$) affects the optical properties of the vacuum around the neutron star which behaves like a dichroic and birefringent medium \citep{adler_photon_1971}. When this vacuum contribution, which is a genuine QED effect, is added to the plasma dielectric and magnetic permeability tensors, it can result in the presence of ``vacuum resonances'' within the neutron star atmosphere.

At the vacuum resonance, the eigenvalues of the wave equation become degenerate, and mode conversion can occur \citep{lai_transfer_2003, ho_ii_2003, ho_iii_2003}. 
Mode conversion at the vacuum resonance can dramatically affect the polarization properties of a highly magnetized atmosphere, and the detection of such properties would be an indication of the presence of vacuum birefringence. Since the physics of this effect is still poorly understood, some  limiting/approximated cases are often assumed in the literature: no mode conversion, complete mode conversion, and partial mode conversion \cite[see][and references therein]{kelly_x-ray_2024-1}.
\begin{figure}
    \centering
    \includegraphics[width=1\linewidth]{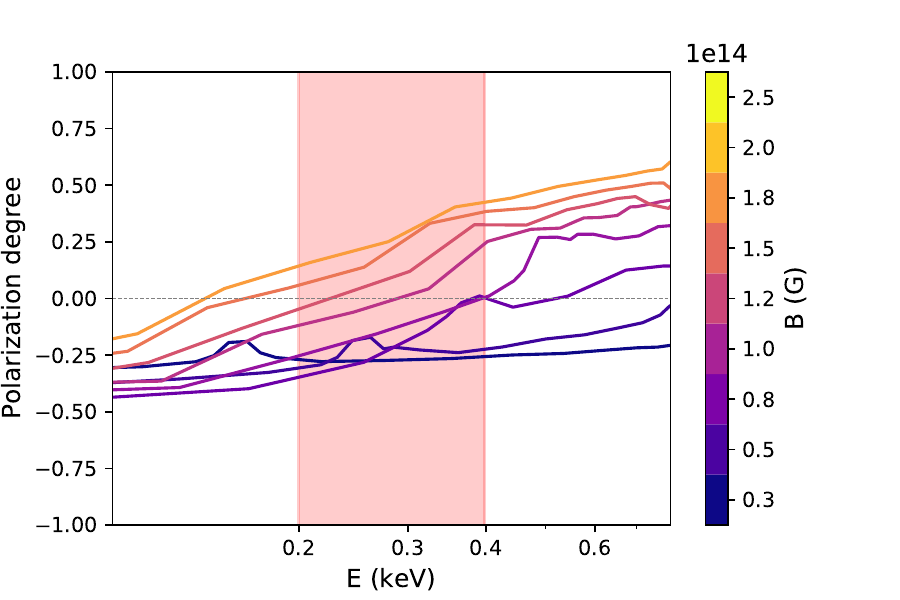}
    \caption{Same as in figure \ref{fig:standard} for atmospheric  models computed accounting for plasma and vacuum contributions and by assuming complete mode conversion at the vacuum resonance.}\label{fig:complete}
\end{figure}
In the limiting case of complete mode conversion, where every photon changes polarization mode, a change in the dominant polarization mode, from O(X) to X(O), is expected to occur, inducing a $90^\circ$ swing in the polarization angle at an energy that increases with decreasing magnetic field strength \cite[]{kelly_x-ray_2024-1}.

In order to illustrate this, in Figure \ref{fig:complete} we show the polarization degree as a function of energy for an atmospheric plasma plus vacuum model computed by assuming complete mode conversion. We find that for magnetic field strengths $\sim 10^{14} \ \mathrm{G}$, the $90^\circ$ angle change (signaled in Figure \ref{fig:complete} by the switch from a negative to a positive value of the degree of polarization) occurs within the observational energy range of REDSoX. We also find that, for larger magnetic field strengths, the effect of complete mode conversion would produce a rapid increase in the degree of polarization with energy in the REDSoX band, from a few $\%$ up to $\sim 45\%$,  while for lower magnetic field strengths the polarization degree remains relatively constant around $\sim 30\%$, and the emission at the source is dominated by the O mode. 

\begin{figure}
    \centering
    \includegraphics[width=1\linewidth]{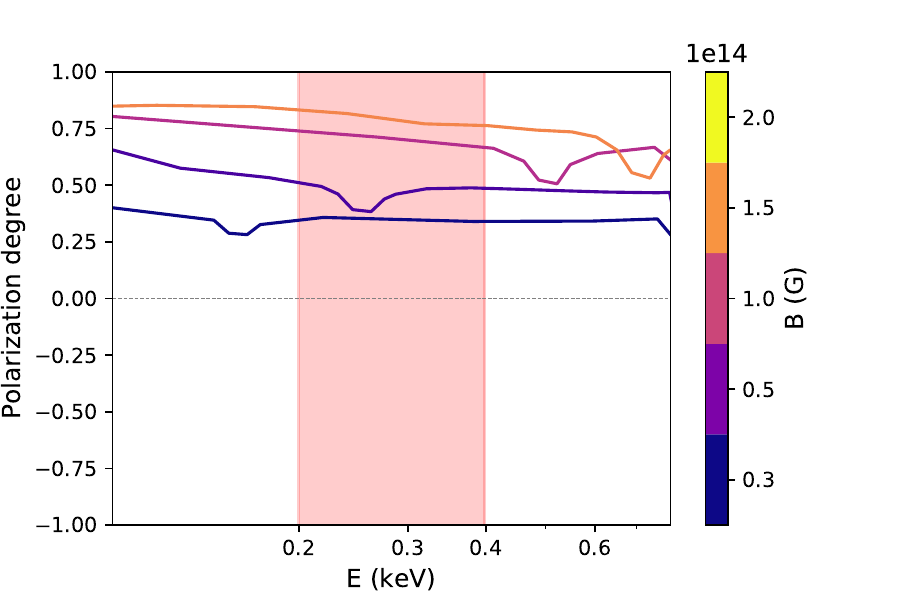}
    \caption{Same as in figure \ref{fig:standard} for atmospheric models computed accounting for plasma and vacuum contributions and assuming partial mode conversion (by using the method as in 
    \citealt{kelly_x-ray_2024-1}, with a 
    probability threshold  $P^\mathrm{th}_\mathrm{con}= 0.1$; see text for details).}
    \label{fig:partial}
\end{figure}

We then investigated the effect of 
partial adiabatic mode conversion at the vacuum resonance, following the formulation by  \citet{lai_transfer_2003, ho_ii_2003}. Specifically, we 
assumed  a conversion probability given by
\begin{equation}
    P_\mathrm{con} = 1 - \exp{\biggl[-\frac{\pi}{2}\biggl(\frac{E_\mathrm{s}}{E_\mathrm{ad}}\biggl)^3\biggl]}\, ,  
\end{equation}
where $E_\mathrm{s}$ is the photon energy at the neutron star surface, and 
$E_\mathrm{ad}$ is the limiting energy for adiabatic mode conversion, which depends on the density of the medium and on the magnetic field strength. We followed the procedure detailed in  \citet[to which we refer for details on the numerical  calculation]{kelly_x-ray_2024-1}, which triggers mode conversion when a photon has a probability larger that a fixed threshold $P^\mathrm{th}_\mathrm{con}$. 
An example of our results  is  shown in Figure \ref{fig:partial}, for $P^\mathrm{th}_\mathrm{con}= 0.1$. 
As can be seen, models computed by assuming partial mode conversion show, in the range below $1 \ \mathrm{keV}$, a polarization degree which is systematically reduced with respect to that expected in the case of a pure plasma atmosphere (see Figure \ref{fig:standard}) and, for instance, can be as low as $\sim 30$--$40\%$ for $B\sim 3 \times 10^{13}$~G. Interestingly, in this case radiation remains X mode dominated across the entire REDSoX energy range because, for this value of the magnetic field strength, the probability criterion is not met for these low-energy photons.  

\section{REDSoX} \label{sec:application}

REDSoX is a NASA-funded sounding rocket mission that conducts X-ray spectropolarimetry for targets such as blazars. The instrument, if operational for a longer duration of time, also has the capability to perform spectropolarimetry for isolated neutron stars, which we explore here.
REDSoX is designed to complement instruments such as IXPE and expand current X-ray polarimetry capabilities below $1\,\mathrm{keV}$.
It is scheduled for launch in late 2027 to observe the blazar Mrk 421. The design utilizes a focusing Wolter I optic, Critical Angle Transmission (CAT) gratings, and laterally graded multilayer mirrors (LGML) at Brewster’s angle to perform spectropolarimetry in the energy range of $0.2$--$0.4\, \mathrm{keV}$ \cite[see][]{marshall_design_2018,marshall_status_2023,garner_current_2024}.

The instrument operates by passing X-rays first through full-shell grazing-incidence Wolter-I focusing optics and then through one of three polarimetry channels consisting of a grating array, LGML, and detector. Focused X-rays pass through high-efficiency dispersive CAT gratings, with zeroth-order light collected by a central imaging CCD detector, used for target acquisition \citep{heine_characterization_2024}. The first-order dispersed light passes through an LGML placed at Brewster's angle $(45^\circ)$ to select for s-polarized light (polarized perpendicular to the plane of incidence) and is then collected by first-order polarimetry CCD detectors \citep{heine_characterization_2024, marshall_design_2018}.

\subsection{Instrument Detection Capability} \label{subsec:method}
In this section we turn to the calculation of the signal emitted by the whole neutron star surface. We explore potential polarization measurements of surface radiation from neutron stars that can be achieved using an instrument such as that on board REDSoX in an orbital space mission. 

We assumed a core-centered dipole magnetic field and divided the north hemisphere of the neutron star surface into six annular patches in magnetic co-latitude, centered at $\theta = (0^\circ, 10^\circ, 30^\circ, 50^\circ, 70^\circ, 89^\circ)$. We adopted a surface temperature profile $T_\mathrm{s} = \max(T_\mathrm{dip}, T_\mathrm{c})$, where $T_\mathrm{dip} =
T_\mathrm{p}|\cos \theta_B|^{1/2}$ represents  the relation between the temperature and the local magnetic field inclination \citep{greenstein_pulselike_1983}, and $T_\mathrm{c}$ is a minimum temperature set to avoid unrealistically low values near the equator; in the following we take $T_\mathrm{p} = 10^7 \ \mathrm{K}$ and $T_\mathrm{c} = 7\times10^{6} \ \mathrm{K}$ \cite[see][for more details]{kelly_x-ray_2024-1}.
Atmospheric models have been produced for each patch, accounting for the inclination of the magnetic field in the radiative transfer calculation \cite[see][for details on the code]{lloyd_model_2003} and using the assumptions discussed in section~\ref{sec:0.1-1keV}. The emission from the south hemisphere is then computed using the same atmospheric models and symmetry arguments.

The observed phase-dependent, monochromatic Stokes parameters depend on the source viewing geometry through the angles $\chi$ and $\xi$ the spin axis makes with the line-of-sight and the
magnetic axis, respectively. Once the viewing geometry is fixed, the simulated specific Stokes parameters
$\mathcal{I_\lambda}, \ \mathcal{Q_\lambda}, \ \mathcal{U_\lambda}$ are computed by summing the contributions of the part of the surface in view at each rotational phase, accounting for general relativistic effects \cite[see][]{page_surface_1996}.  To calculate the emission at a distant observer, we use the ray tracing code described in \citet{zane_unveiling_2006}, \citet{taverna_polarization_2015} and \citet[see also \citealt{kelly_x-ray_2024-1, kelly_x-ray_2024}]{gonzalez_caniulef_polarized_2016}. 

In this section, we present results for $\chi=80^\circ,\xi = 0.05^\circ$, which is a viewing geometry chosen among the most promising configurations, i.e. those for which the observed polarization from the source is the largest \citep{gonzalez_caniulef_polarized_2016}.

The effect of interstellar absorption is taken into account according to the \texttt{tbabs} model
\footnote{\url{https://pulsar.sternwarte.uni-erlangen.de/wilms/research/tbabs/}.}, assuming the elemental abundances from \cite{wilms_absorption_2000} and the cross sections from \cite{verner_atomic_1996}.
We used a fiducial hydrogen column density
$N_{\rm{H}} = 1.5\times10^{20} \ \mathrm{cm}^{-2}$, typical of isolated neutron stars.

For the calculation of the polarization measured by the detector,
the Stokes parameters $\mathcal{I}_{\lambda},\, \mathcal{Q}_{\lambda}, \, \mathcal{U}_{\lambda}$, obtained by the code,
are convolved with the absorption and then rescaled 
with the normalization factor $N$,
which is obtained from
\begin{align}
    N \int_{\lambda_0}^{\lambda_1} \mathcal{I}_{\lambda} \ \diff{\lambda} = 1 \, \mathrm{count/cm^2/s}\,,
\end{align}
where $\lambda_0,\,\lambda_1$ are the wavelengths (in Ångström, Å) corresponding to the energy bandpass $0.4, 0.2 \ \rm{keV}$; the rescaled quantities are denoted by $I_\lambda, \, Q_\lambda, \, U_\lambda$.
For an observation with total counts $C_{\rm{obs}}$ throughout its duration,
$I_{\lambda}, \ Q_{\lambda}, \ U_{\lambda}$ can be rescaled again (at every wavelength)
by the factor $C_{\rm{obs}}$
to obtain the specific quantities.

The expected generalized count rate $R_{\lambda,i}$ per wavelength bin for each detector ($i = 1,\,2,\,3$) is computed by convolving the normalized Stokes parameter values with the REDSoX instrument response such that
\begin{align}  \label{eqn:count_rate}
    R_{\lambda,i} = A_\lambda I_\lambda +A_\lambda\mu_\lambda[Q_\lambda \cos(2\theta_i)+U_\lambda \sin(2\theta_i)].
\end{align}
Here, $\mu_\lambda$ is the modulation factor and $\theta_i$ is the orientation of the detector $i$ 
with respect to the $Q$ and $U$ reference frame, i.e. $-Q$ is aligned with $\theta=0^\circ$ 
\cite[the three detectors are at $120^\circ$ from each other][]{marshall_design_2018}. In the following, we use $\theta_1=0^\circ$, $\theta_2=120^\circ$ and $\theta_3=240^\circ$. 
The integrated area of the instrument, given by $\int A_\lambda d\lambda = 185 \mathrm{cm}$Å, was taken from the REDSoX design paper \cite{marshall_design_2018}, where the effective area $A_\lambda$ was taken from their Fig.~15.

The measured Stokes parameters per wavelength bin can then be obtained from the detector count rates $R_{\lambda,i}$ by
\begin{align}
    I_{\lambda,\rm{det}} &= (R_{\lambda,1} + R_{\lambda,2} + R_{\lambda,3}) / (3A_\lambda) \nonumber \\
    Q_{\lambda,\rm{det}} &= (2R_{\lambda,1} - R_{\lambda,2} - R_{\lambda,3})/(3\mu A_\lambda) \nonumber \\
    U_{\lambda,\rm{det}} &= (R_{\lambda,2} - R_{\lambda,3})/(3^{1/2}\mu A_\lambda)
\end{align}
from which follow the polarization degree  $\mathrm{PD}=\sqrt{Q_{\rm{det}}^2 + U_{\rm{det}}^2}/I_{\rm{det}}$ and angle $\mathrm{PA}=\arctan(U_{\rm{det}}/Q_{\rm{det}})/2$.

The instrument has the spectral resolution (and line response function, LRF) of an objective grating spectrometer.  The LRF of a grating spectrometer is typically described as a Gaussian with a FWHM of $\delta\lambda = P \delta x/z$, where $P$ is the grating period, $\delta x$ is the FWHM of the instrument along the dispersion, and $z$ is the distance of the grating from the focal plane; for REDSoX $P = 2000$ \AA\ \citep{marshall_design_2018} and $\delta x = 0.12$ mm \citep{marshall_rocket_2024}.  REDSox has an unusual placement of gratings but most are placed at distances within 10\% of the on-axis case: $z = P / (\sqrt{2} G)$, where $G = 0.88$ \AA/mm is the gradient of the multilayer spacing \citep{marshall_rocket_2024}, so $z$ is 1600 mm ($\pm 10$\%).  Thus, $\delta \lambda$ is within 10\% of 0.15 \AA.  The spectral bins used in this paper are all much larger than $\delta \lambda$, so the count rates $R_{\lambda,i}$ are statistically independent of each other.

The minimum detectable polarization (MDP) of the instrument at the $99\%$ confidence level is defined as
\begin{align}
    \mathrm{MDP}=\frac{4.292}{\sqrt{\sum_i\left(\mu_i^2 C_i^2 \right)}}\,,
    \label{mdp}
\end{align}
where the sum is extended over all the wavelength bins and $\mu_i$, $C_i$ are the instrument modulation factor and source count rate in the $i$-th wavelength bin.

We estimate that background will be negligible.  The particle background for REDSoX was estimated to be 0.001 count/s \cite[][section 3.5]{marshall_design_2018}.  We use a sky background estimate of 540 ph/cm$^2$/s/sr/keV \citep{marshall_design_2018}, or 3.9 ph/cm$^2$/s/sr/\AA\ at 0.3 keV.  For a solid angle of 2.5 arcminutes radius, where the instrument sensitivity drops to 50\% of peak \cite[][Fig. 16]{marshall_design_2018}, and the integrated area of REDSoX (185 cm$^2$\AA), the sky background should contribute about 0.0012 cts/s.  Thus, the total background is estimated to be about 0.0022 cts/s, which is negligible compared to the expected count rates of the targets (section \ref{sec:casestudies}).

\subsection{Results and statistics}
To illustrate the possible features from isolated neutron stars,
we computed mock spectra for two atmospheric models, (a) pure-plasma atmosphere (i.e. without vacuum corrections), with $B = 5\times 10^{13} \ \mathrm{G}$, and (b) plasma-plus-vacuum atmosphere with complete mode conversion and $B = 10^{14} \ \mathrm{G}$. 
We also computed the case of a plasma plus vacuum model with no mode conversion and obtained quantitatively similar results as in case (a); therefore scenario (a) can be taken as representative of both these situations. 
These magnetic field strengths were chosen such that a proton cyclotron feature (model a) or a $90^\circ$ polarization angle swing (b) are present in the instrument detectable energy range.

To create a mock spectrum, we realized a Poisson-distributed count number $n_{\lambda, i} \sim \mathrm{Pois}(k_\lambda)$ in each detector channel $(i = 1, 2, 3)$, where $\lambda$ denotes the channel wavelength. The mean $k_\lambda$ is the expected number of photons in each channel, which is obtained by multiplying the model spectrum count rates, given by equation \eqref{eqn:count_rate}, with the observation time.

\begin{figure}
    \centering
    \subfigure[]{\includegraphics[width=1\linewidth]{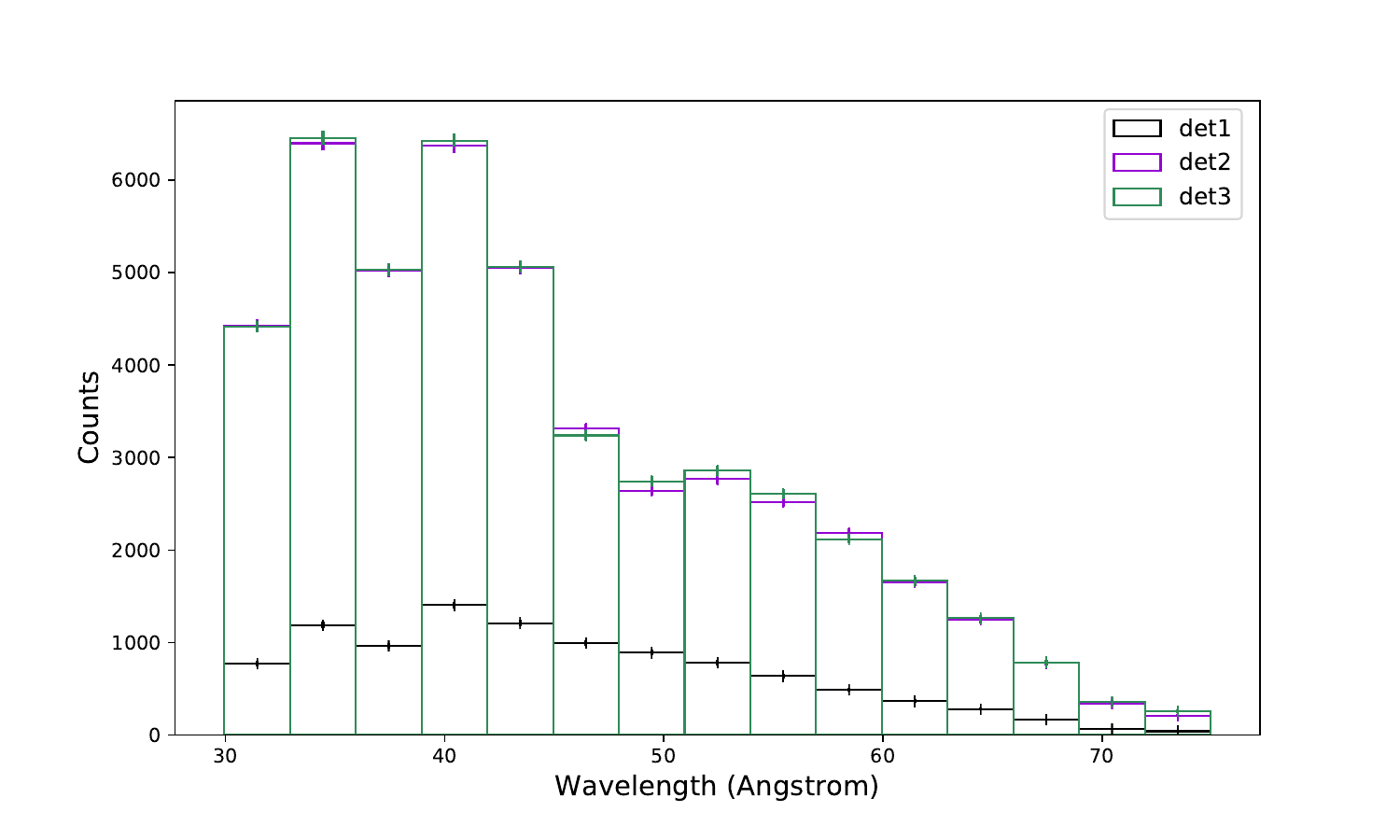}}
    \subfigure[]{\includegraphics[width=1\linewidth]{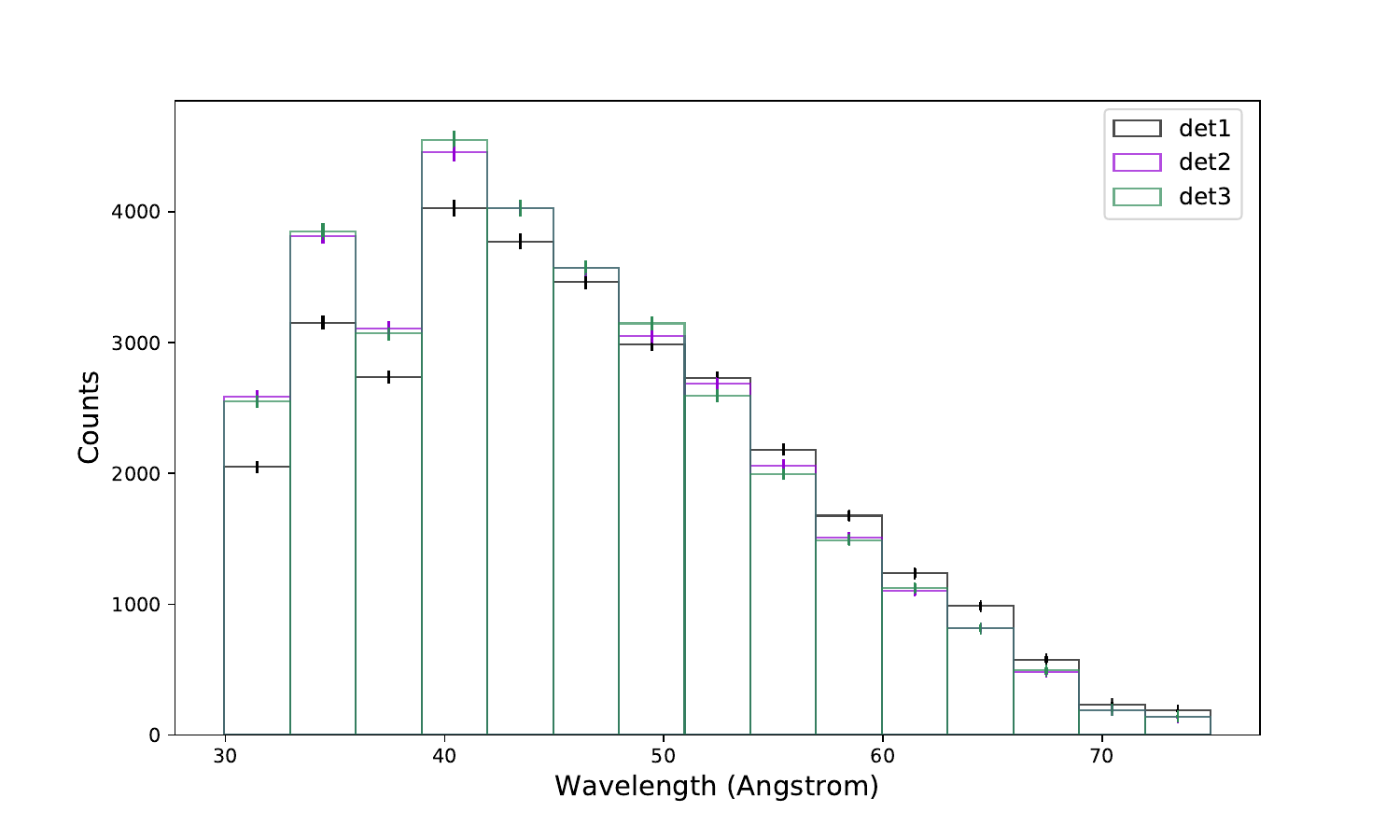}}
    \caption{Counts (as a function of wavelength) for each detector channel (det1, det2, det3) calculated assuming emission from an atmosphere covering the whole surface in the pure plasma (a; upper panel) and plasma plus vacuum with complete mode conversion (b; lower panel) scenarios. The  dipole magnetic field strengths are $B = 5\times 10^{13}$ and $10^{14} \  \mathrm{G}$ in the two cases, respectively. A total of  $100,000$ counts is assumed to be collected across all three detectors.}
    \label{fig:mock_spectra}
\end{figure}

Figure~\ref{fig:mock_spectra} shows an example of the mock counts in each detector versus the wavelength, assuming a bin size of $\Delta\lambda = 3$ Å and a total of $ \sim 100,000$ counts in all three detectors. The first dip, present in both cases at $\sim 37$ Å, is a feature of the detector response. The second dip which can be seen at $\sim 50$ Å ($\sim 0.25 \ \mathrm{keV}$) in detectors 2 and 3 in Figure \ref{fig:mock_spectra}(a), is due to the proton cyclotron feature. In this example,  detector 1 collects a significantly reduced number of counts compared to detectors 2 and 3: this is due to the orientation of the detectors with respect to the reference frame of $Q$ and $U$, together with the fact that, for the parameters of this model, there is a large difference between the expected values of $Q$ and $U$. In fact, from equation \ref{eqn:count_rate}, when $\theta_1 = 0^\circ$, if $Q$ is negative and $|Q|\gg |U|$ (as in this case), detector 1 will collect much fewer counts than the other two detectors. This difference in the detector count rate does not appear in the complete mode conversion scenario shown in  Figure \ref{fig:mock_spectra}(b). 

Additionally, as expected due to the higher magnetic field strength in scenario (b), no second dip (corresponding to the proton cyclotron line) is visible, and the reduction in counts due to detector response and blackbody flux of the source is much more gradual.

\begin{figure}
    \centering
    \includegraphics[width=1\linewidth]{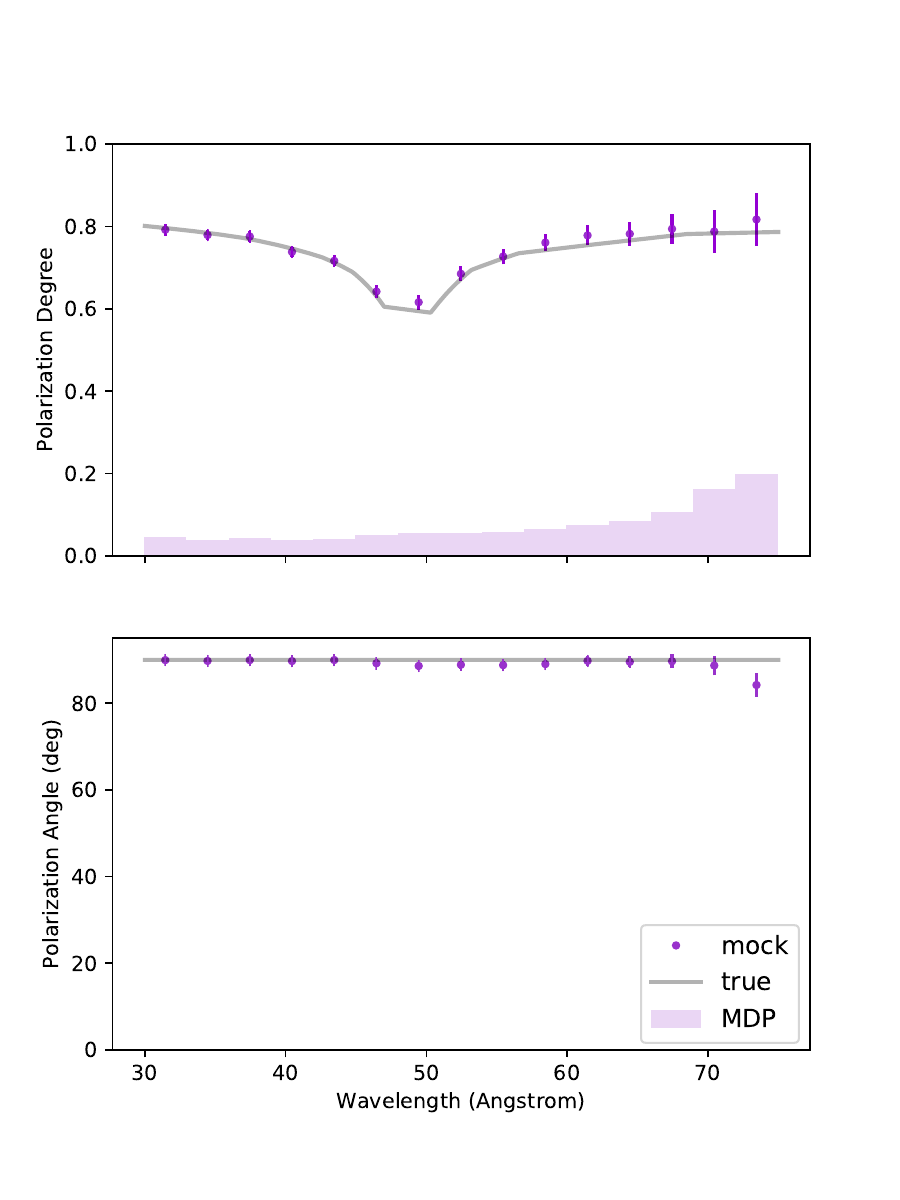}
    \caption{Polarization degree (upper panel) and angle (lower panel) as a function of the wavelength for scenario (a). The model from which the mock data were produced is shown by the solid line. Here, $B = 5\times 10^{13}\,\mathrm{G}$. 
    }
    \label{fig:polarisation_proton_cyc}
\end{figure}

Figure~\ref{fig:polarisation_proton_cyc} shows the energy(wavelength)-resolved polarization degree and angle for the scenario (a). As can be seen, in this case a high polarization can be detected,  while the $\mathrm{PA}$ remains constant across the entire energy range of the detector. The  energy-integrated values in the REDSoX band are $\mathrm{PD} = 74.5 \pm 0.5 \%$ and $\mathrm{PA} = 89.6^\circ \pm 0.2 ^\circ$. 
The dip in polarization due to proton cyclotron resonance is clearly visible at $\sim 50$Å ($\sim 0.25 \ \mathrm{keV}$), where the polarization degree decreases from $\sim 80\%$ to below $60\%$. 
The polarization is well above the MDP across the entire energy range.

To investigate how many total counts are required to ensure that $99\%$ of possible observations will give a statistically significant detection of a similar proton cyclotron feature, we simulated sets of $10,000$ mock spectra for different values of the total counts.
We computed the $\chi^2$-statistic for each mock spectrum (with $5$ Å binning)
using a constant polarization model (i.e. fitting the mock data with a horizontal straight line).
A reduced $\chi^2$ of $2.5$ (with $8$ degrees of freedom) was chosen as our goodness-of-fit criterion: this means that those mock data realizations that result in a higher $\chi^2$ value statistically rule out a constant polarization model at the $98\%$ confidence level.

We found that for the set of mock spectra with $25,000$ total counts, a constant polarization model provides an unacceptable fit ($\chi^2_\mathrm{red} > 2.5$) for more than $99$\% of the realizations in the set.
With $15,000$ total counts, instead, 
only about $77$\% of the mock spectra can rule out a constant polarization model at the same confidence level.
However, when the total counts exceed $5,000$,
all the mock spectra in the set are better fit by the atmospheric model with the proton cyclotron line than by a model with constant degree of polarization.

\begin{figure}
    \centering
    \includegraphics[width=1\linewidth]{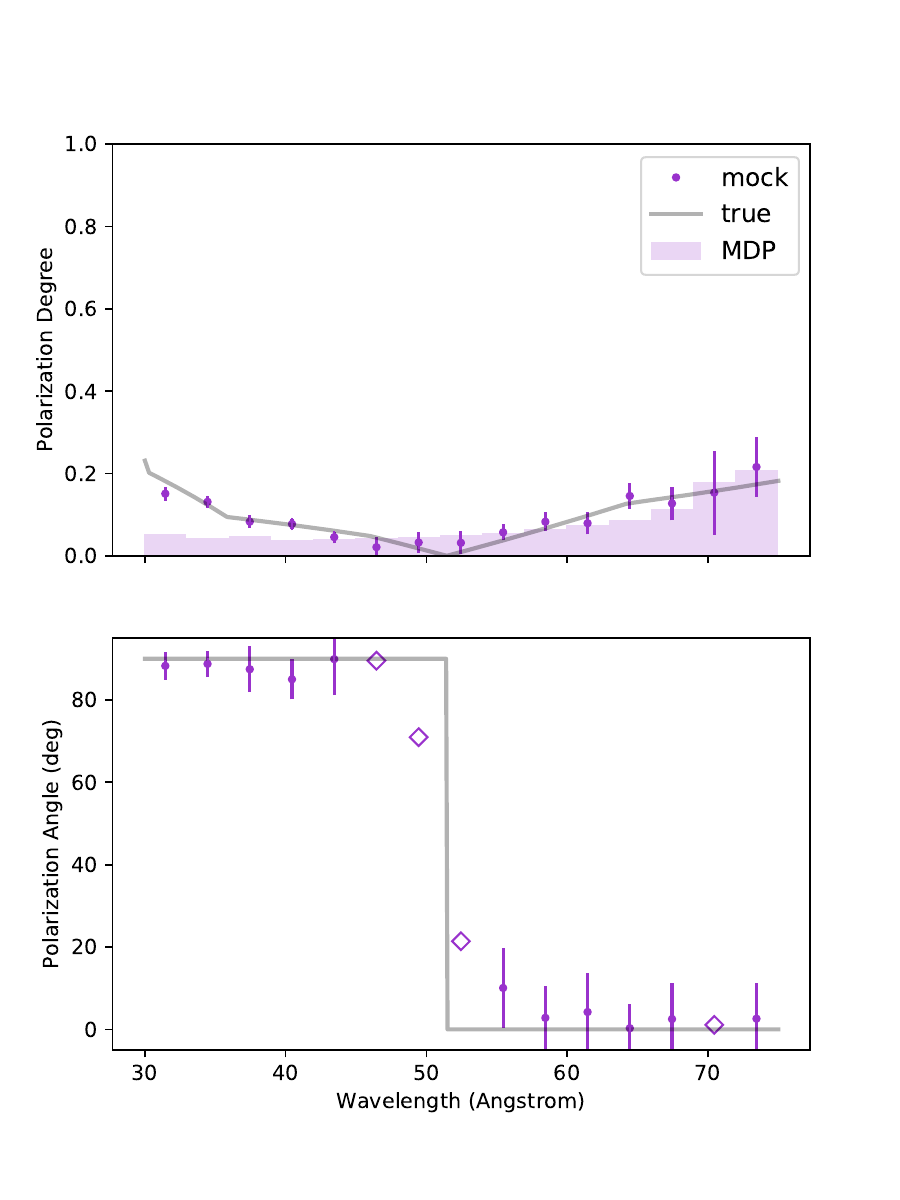}
    \caption{Same as in figure \ref{fig:polarisation_proton_cyc} for model (b). Here, $B = 10^{14}\, \mathrm{G}$. When the polarization degree is below MDP, the corresponding error bars in $\mathrm{PD}$ are shown at $90\%$ c.l. and the diamonds indicate an unconstrained polarization angle.
    }
    \label{fig:polarisation_completemode}
\end{figure}

In scenario (b), corresponding to a plasma plus vacuum model with complete mode conversion and a larger magnetic field, a significantly lower overall polarization is expected  when compared with model (a). The energy-integrated value (in the REDSoX band) is $\mathrm{PD} = 3.5 \pm 0.5 \%$, while the corresponding MDP is $1.4\%$.
As shown in Figure \ref{fig:polarisation_completemode}, the polarization changes significantly with energy in the REDSoX range:  $\mathrm{PD}$ reduces from almost $20\%$ to $\approx 0 \%$ at $50$Å ($\sim 0.25 \ \mathrm{keV}$) before increasing again to $\sim20\%$ at $70$ Å.
At the wavelength where $\mathrm{PD}\approx 0 \%$, the polarization angle swings by $90^\circ$, indicating a change in the dominant polarization mode induced by complete mode conversion at the QED vacuum resonance.
For the majority of the bins in the REDSoX wavelength band, except for those near the dominant mode switch and for the high wavelengths, the predicted polarization is above MDP. For bins with polarization below MDP, the polarization angle is unconstrained (these points are shown as diamonds in the figure).

We then tested whether the $90^\circ$ swing
in $\mathrm{PA}$ could be detected, again using sets of $10,000$ spectral realizations characterized by a different number of total counts.  We computed the $\chi^2$ statistic considering a model with a linear variation of the polarization angle (i.e. a fit with a non horizontal straight line) across the energy range, binning data to $5$ Å for $7$ degrees of freedom.
Our simulations showed that, for spectra with more than $15,000$ total counts, a model with a constant rate of change in the polarization angle was statistically ruled out ($\chi^2_{\mathrm{red}} > 2.5$) for $>99\%$ of the set.
However, for $15,000$ counts, the majority of the wavelength bins are below the MDP, and as such the polarization angles are unconstrained. In simulations with a low number of counts, the results are less statistically significant because there are fewer data points for fitting. It is only for counts $> 25,000$ that the polarization degree is above MDP in the majority of wavelength bins, when binning to $5$ Å. If we increase the bin size to $\sim 8$ Å, the majority of the spectral bins for $\gtrsim 15,000$ counts are above MDP.

\section{Case Studies} \label{sec:casestudies}
\subsection{RX J1856.5-3754} \label{sec:1856}
RX~J1856 is the brightest and closest XDINS,
with a distance from Earth of $\sim 120 \ \mathrm{pc}$ \citep{walter_revisiting_2010}. It has a rotational period of $P\sim 7 \ \mathrm{s}$ and a very low pulsed fraction of $\sim 1.3 \%$ \citep{tiengo_xmm-newton_2007}.
The X-ray spectrum ($0.1$--$10\, \mathrm{keV}$) is well fit by two blackbody components \cite[with temperatures $T_\mathrm{h} \approx 60 \ \mathrm{eV}$ and $T_\mathrm{c} \approx 40 \ \mathrm{eV}$, respectively, see][]{sartore_spectral_2012}. 
The source has also been observed in the optical band, with a sizable degree of polarization ($\approx 16 \%$), strongly suggesting that QED vacuum birefringence is at work around the star \citep{mignani_evidence_2017}.

For our simulations, we follow the model described in \cite{gonzalez_caniulef_polarized_2016}, assuming a dipole magnetic field with strength $B_\mathrm{p} = 10^{13} \ \mathrm{G}$ at the pole, a polar temperature $T_\mathrm{p}=60\,\mathrm{eV}$ and an equatorial temperature $T_\mathrm{e}=40\,\mathrm{eV}$, both measured at the observer, and $N_{\rm{H}} = 0.5\times10^{20} \ \mathrm{cm}^{-2}$.
We again divide the surface into six annular patches with magnetic co-latitude centered on $\theta = (0^\circ, 10^\circ, 30^\circ, 50^\circ, 70^\circ, 89^\circ)$ in the north hemisphere. As previously, integration in the south hemisphere is then performed using symmetry properties. 

We investigated the cases of pure-plasma and plasma plus vacuum atmospheric emission models, the latter computed either assuming  no mode or  complete mode conversion. Again, we found that the first two cases  give the same results in the REDSoX band, therefore they will not be discussed separately. We also considered the case of condensed surface emission, with the emissivity computed following 
\cite[][see also \citealt{gonzalez_caniulef_polarized_2016, taverna_x-ray_2020}]{potekhin_radiative_2012} by assuming iron chemical composition.  As discussed in \citet{mignani_evidence_2017}, in the limit of free-ions the emission expected from a condensed iron layer is not compatible with the optical polarization properties of the source; therefore this limit  will not be considered in our simulations. 

We performed simulations by accounting for or turning off QED vacuum birefringence effects in the photon propagation in the vacuum from the star surface to the observer; these models will be referred to as QED on and QED off in the following. The viewing geometry is set at $\chi, \xi = (14^\circ.0, 3^\circ.0)$ for all the atmospheric emission models\footnote{We caveat that in 
\citet{mignani_evidence_2017} the viewing geometry has been derived by considering those models that predict a value of the optical polarization in agreement with observations. Since the assumptions on some of the atmospheric emission model are slightly different in this work, the constraint on the viewing geometry may be slightly impacted. This is not explored in this work.} and $\chi, \xi = (21^\circ.7, 5^\circ.5)$ for the condensed surface model in the limit of fixed-ions, according to the constraints on the viewing geometry derived by \citet[see their Table 1]{mignani_evidence_2017}.
\begin{figure*}
    \centering
    \includegraphics[width=1\linewidth]{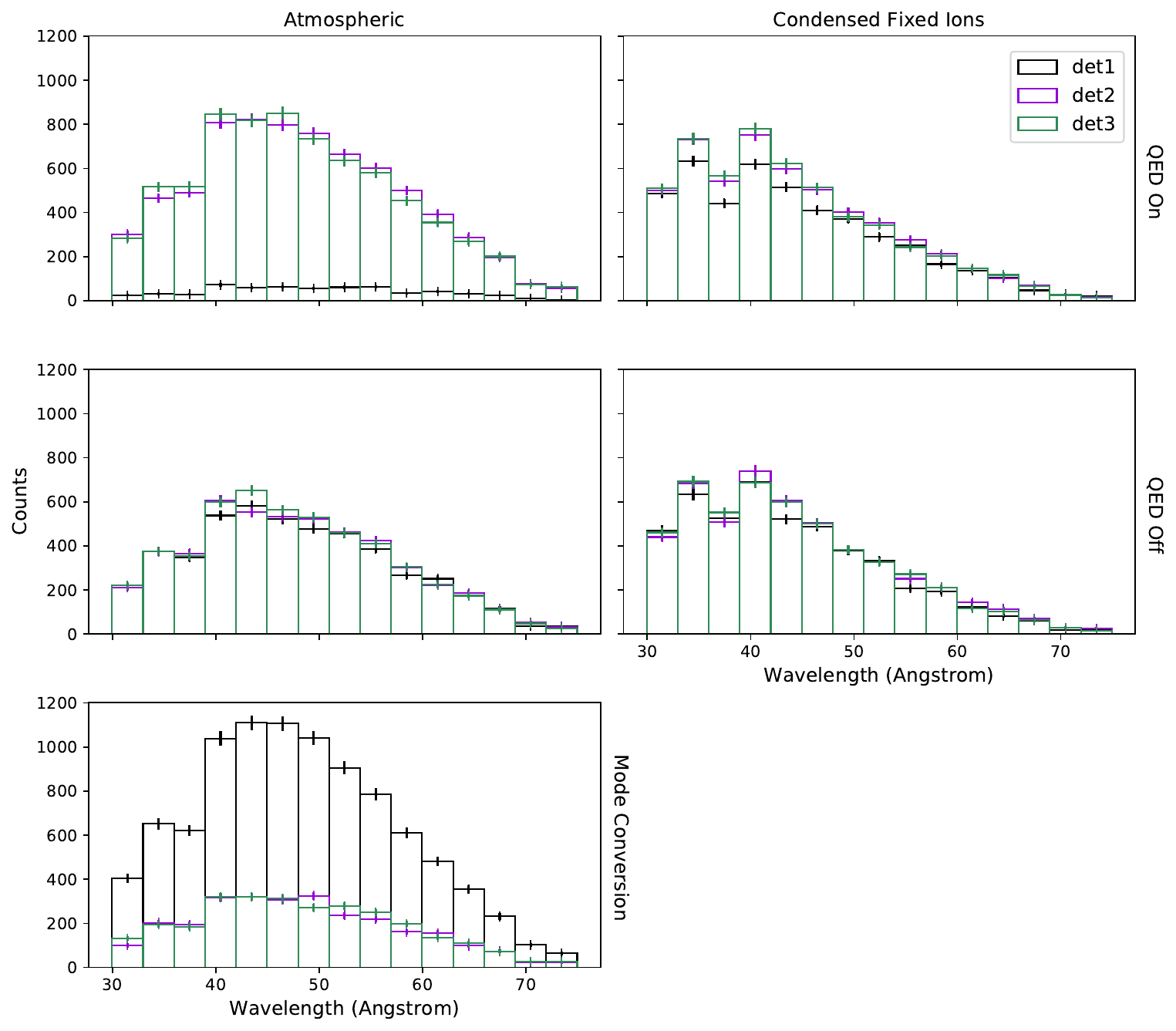}
    \caption{Expected counts for each instrument detector channel (det 1, 2, 3),  binned at  $3$ Å,   calculated from a mock spectrum of RX~J1856 assuming an observation time of 
    125 ks and a count rate of 0.12 cts/s. All models assume
    emission from the whole surface. The first two rows show the case of 
    a plasma plus vacuum atmosphere (no mode conversion, left column) and a condensed  iron surface (fixed-ions models, right column); both models are computed by turning on or off effects of QED vacuum birefringence in the photon propagation in the vacuum from the star to the observer (first and second line, respectively). The third row shows the case of a plasma plus vacuum  atmospheric model with complete mode conversion and QED effects in the photon propagation outside the star.}
    \label{fig:mock_spectra_1856}
\end{figure*}
We assume a count rate of $0.12 \, \mathrm{cts/s}$ and an observation time of 125 ks, which give a total of $\sim15,000$ expected counts over the three detectors, and use a bin size of $3$ Å.

Figure \ref{fig:mock_spectra_1856} shows the detector counts over the REDSoX wavelength band for the various models. 
In all cases, we see no significant features in the detected counts per bin; for a field strength of $10^{13}\, \mathrm{G}$ the proton cyclotron feature falls outside the REDSoX energy range. 

\begin{figure*}
    \centering
    {\includegraphics[width=1\linewidth]{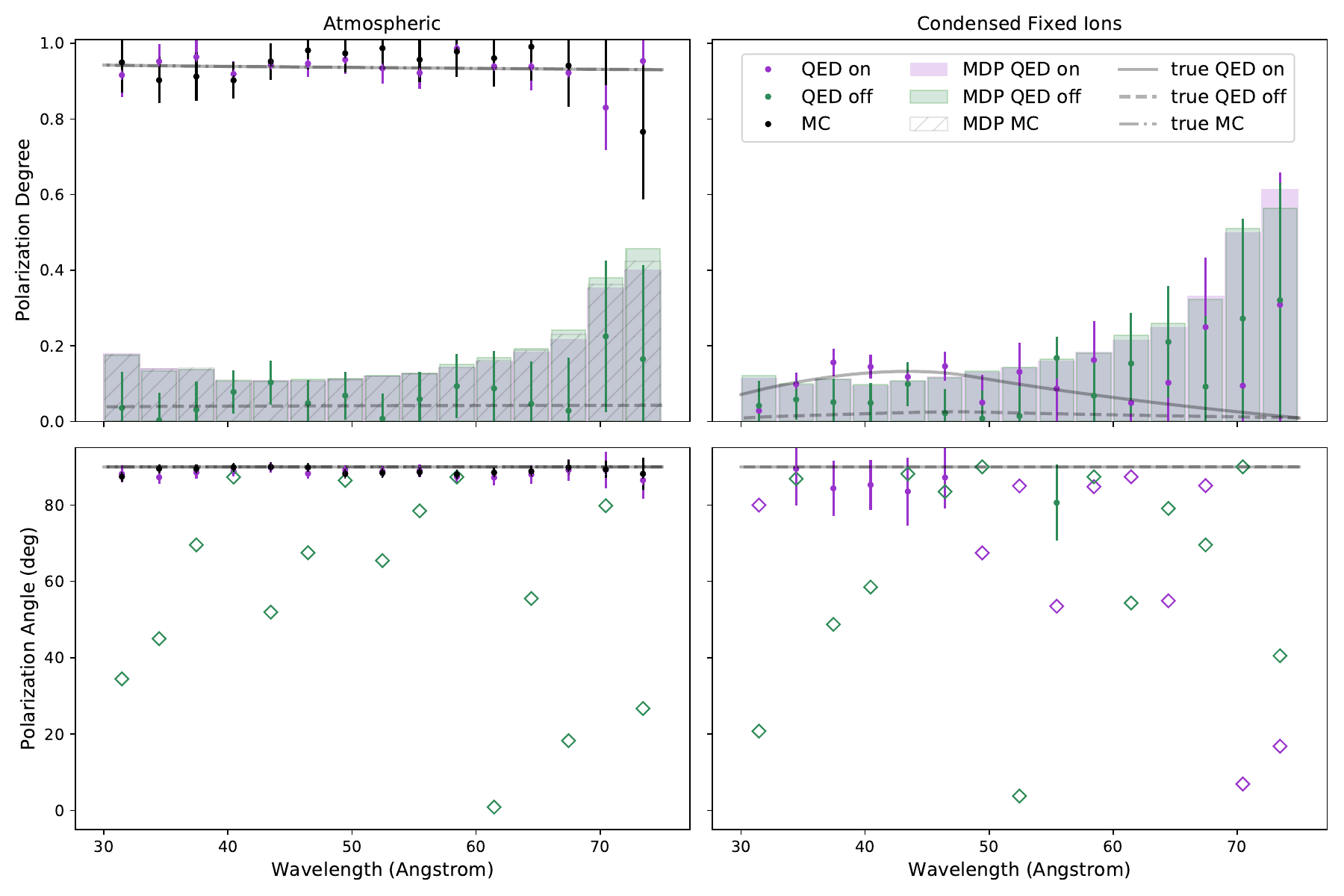}}
    \caption{Expected polarization degree (top row) and angle (bottom row) of RX~J1856  as function of the wavelength, for different models: a plasma plus vacuum atmospheric emission model with complete  conversion ('MC'),  a pure plasma (no vacuum corrections in the atmosphere) atmospheric emission model (both computed for $\chi=13^\circ$, $\xi = 4^\circ$), a condensed surface emission model computed in the fixed ion limit ($\chi = 21.7^\circ$, $\xi =  5.5^\circ$).  Results  were obtained with and without the inclusion of  QED vacuum effects  in the photon propagation outside the star (``QED on/off''). An observation time of 
    125 ks and a count rate of 0.12 cts/s are assumed. The figure shows  the mock data with their $1 \sigma$ errors (points), the original synthetic polarization spectra (full lines) and the MDP for each simulated observation (shaded area). When the polarization degree  is below MDP, the error bars  are shown at $90\%$ confidence level. The diamonds represent an unconstrained polarization angle.}
    \label{fig:polarization_1856}
\end{figure*}

The corresponding polarization features are illustrated in Figure~\ref{fig:polarization_1856}. First, we considered models computed by accounting for vacuum QED effects in the photon propagation from the star to the observer (``QED on'' models in the figure). As expected, in this case we found that in the case of plasma plus vacuum atmospheric emission without mode conversion radiation is highly polarized (wavelength-integrated $\mathrm{PD} = 94.6 \pm 1.2\, \%$ and $\mathrm{PA} = 89^\circ.98\pm 0^\circ.45 $ across the entire REDSoX energy band). This value of $\mathrm{PD}$ is significantly higher than the energy-averaged MDP for this observation, which is  $4.4\%$.
The complete mode conversion model shows very similar features  ($\mathrm{PD} = 93.9 \pm 1.6\, \%$ and $\mathrm{PA} = 85^\circ.80 \pm 2^\circ.7 $).
This is not unexpected since, for a field $B=10^{13}\ \mathrm{G}$ and the temperature values used here, the vacuum resonance density falls outside the X and O mode photospheres, so that mode conversion has little effect on the overall polarization properties of the emergent signal \cite[see][]{kelly_x-ray_2024-1}.
We also tested an atmospheric model computed assuming partial mode conversion for different values of the threshold probability $P_\mathrm{con}^\mathrm{th}$, and found a similarly high emergent polarization.  
In comparison, if QED effects related to photon propagation in the vacuum outside the atmosphere are not included (``QED off'' cases),  the polarization is significantly reduced (energy-integrated $\mathrm{PD} = 3.7 \pm 1.6\, \%$ for a pure plasma atmosphere). 
A similar trend is found when considering emission from a condensed surface, in which case $\mathrm{PD} = 13.6 \pm 1.2\,\%$ in the QED on case and only $3.2 \pm 1.2\,\%$, in the QED off one.

For both the QED on atmospheric models presented here, the observed polarization degree turns out to be well above MDP in all wavelength bins (see Figure~\ref{fig:polarization_1856}). 
However, for the QED off atmospheric model and for both the QED on and QED off condensed surface models, 
the observed  polarization is below MDP in almost all of the wavelength bins. This results in unconstrained polarization angles (indicated by the diamond symbol in Figure \ref{fig:polarization_1856}, $90\%$ error regions are shown for the corresponding detected $\mathrm{PD}$.
However, the QED on the condensed surface model does produce some significant detections in a few wavelength bins (around $40$ ), and it can be observed to be different from the QED off case.

In summary, an observation of this source lasting 125 ks should, at the very least, allow us to test the structure of the outer layers, and, if the emission is atmospheric, 
to detect a degree of polarization of almost $100\%$ with high significance, 
confirming that QED vacuum birefringence is at work.

\subsection{RX J0720.4-3125} \label{sec:0720}

The second brightest XDINS, RX J0720, shows two absorption features in the spectrum \cite[at $\sim 270 \ \mathrm{eV}$ and $\sim 750 \ \mathrm{eV}$, respectively,][]{haberl_phase-dependent_2004, borghese_discovery_2015}, and is a very interesting candidate for soft X-ray polarimetry.
The first, broad ($\Delta E \sim 70 \, \mathrm{eV}$) absorption feature, is present at all rotational phases. If it is a proton cyclotron line, the feature should be related to the global field and implies a dipolar magnetic field strength $B_\mathrm{dip} = 2.5\times10^{13} \ \mathrm{G}$ at the equator of the neutron star.
Absorption has also been speculated to be associated with an atomic transition \citep{haberl_evidence_2006}.
The second absorption feature at $\sim 750 \ \mathrm{eV}$ is instead frequency-dependent and peaks in the $0.1$--$0.3$ phase interval \cite[see Table 2 in][]{borghese_discovery_2015}. As discussed by these authors, it is possible that it is produced by proton cyclotron scattering/absorption in a local magnetic loop with $B\sim 2\times10^{14} \ \mathrm{G}$ above the emitting cap.
This is particularly tantalizing. Even if the second feature is outside the REDSoX energy band, the fact that it hints at the presence of strong magnetic field components near some parts of the star surface may open the possibility of using this instrument to detect features caused by the atmospheric QED vacuum resonance (such as the $90^\circ$ swing in $\mathrm{PA}$ discussed previously). 

We assumed an observation time of 200 ks and a source count rate of $0.06 \, \mathrm{cts/s}$
, for a total of $12,000$ counts across the whole observation. We performed  simulations mimicking the scenario  by \citet{borghese_discovery_2015} and assumed a thermally emitting cap with radius $R_\mathrm{BB} \approx  6 \ \mathrm{km}$ and temperature $kT_\mathrm{BB} \approx 120 \, \mathrm{eV}$\footnote{Our choice of a slightly higher temperature than that reported in \citet{borghese_discovery_2015}, $kT_\mathrm{BB} \approx 80 \, \mathrm{eV}$, is dictated by numerical convenience. Although this will affect the overall spectrum, we expect that it only marginally affects the polarization results \cite[see][]{kelly_x-ray_2024-1}.}. We assume $N_{\rm{H}} = 1.5\times10^{20} \ \mathrm{cm}^{-2}$.
Once again, we consider plasma plus vacuum atmospheric models computed assuming no or complete mode conversion, and a condensed surface emission model, this time exploring both the fixed- and free-ion limits. Again, we derived the observables either with or without accounting for QED vacuum birefringence effects in the photon propagation outside the star. 

Our emission cap is centered at the pole, and extends in colatitude up to $\approx 30^\circ $. In order to account for the varying dipolar field across the cap, we divide the cap into three annular patches at $\theta = (0^\circ, \,15^\circ,\, 30^\circ)$, and calculate the  emission for each of them, assuming $B_\mathrm{p} = 5\times10^{13}\, \mathrm{G}$ at the pole. 
\cite{borghese_discovery_2015} invoked the presence of a small-scale highly magnetic structure (a magnetic loop) to explain the phase-dependent feature observed (off peak in phase)  at $\sim 720\ \mathrm{eV}$ in terms of proton cyclotron absorption. Here 
this higher-$B$ component is simulated by the addition of a fourth patch
with $B=2\times10^{14} \, \mathrm{G}$ and $\theta_\mathrm{B}=0^\circ$, which extends in a limited surface area, and is offset with respect to the pole. We explored different sizes for this patch, finding similar results; in the following we report an example in which it extends   
in the range $10^\circ < \theta< 30^\circ$, and $0^\circ < \phi < 40^\circ$.

\begin{figure*}
    \centering
    \includegraphics[width=1\linewidth]{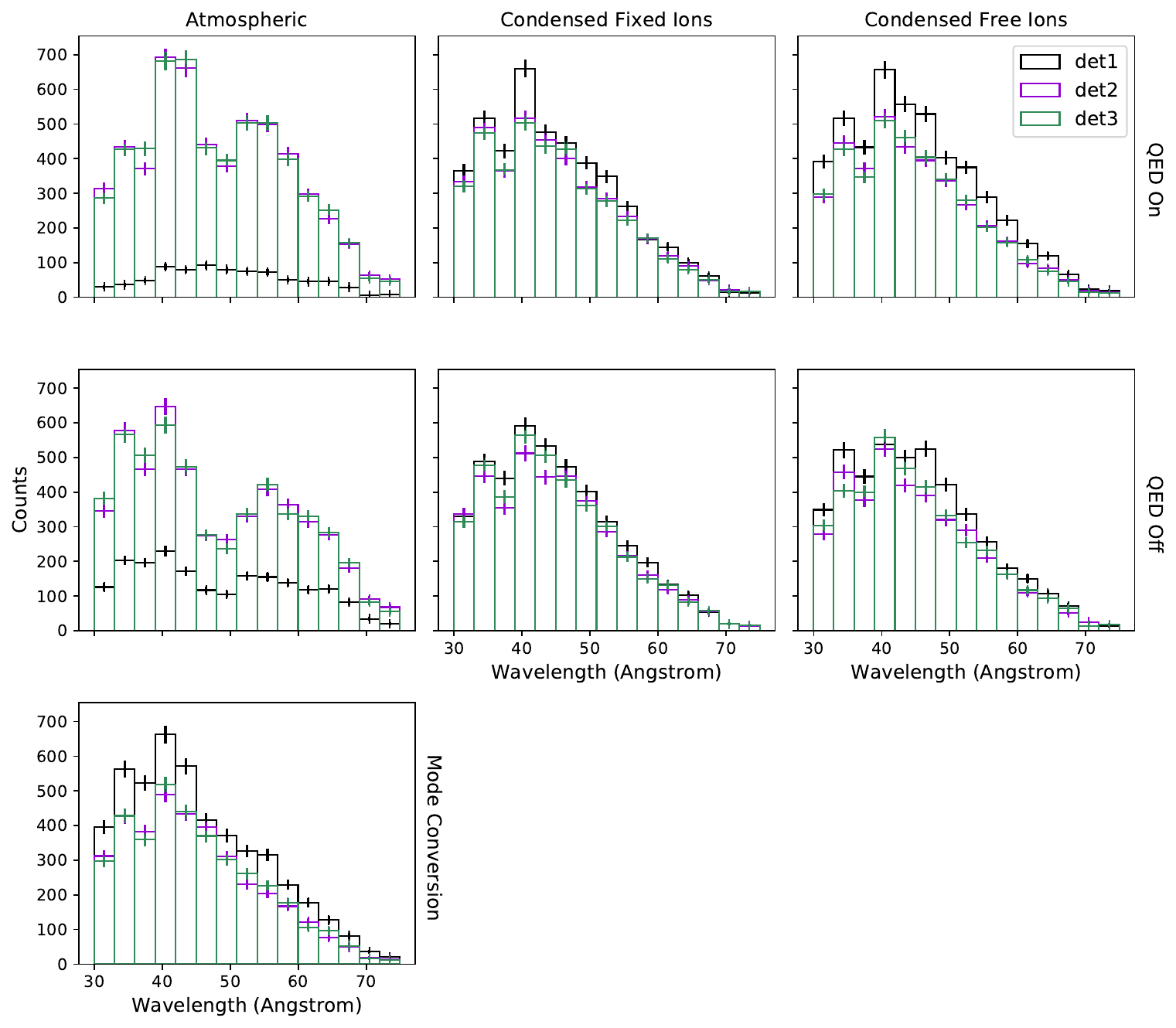}
    \caption{Counts, binned to $3$Å, for each instrument detector channel (det 1,2,3) calculated from a mock spectra of RX~J0720 assuming an observation time of 200 ks and a count rate of $0.06 \, \mathrm{cts/s}$. Spectra are produced for different cases. First row: a plasma plus vacuum atmosphere model with no mode conversion (left) and a condensed iron surface emission model  in the limit of fixed ions (center) and free ions (right), all 
    ``QED on". Second row: a pure plasma atmosphere model (left) and the same condensed surface models as in the first row, but ``QED off''. Third row: an atmospheric model computed assuming complete mode conversion and ``QED on''.  The viewing geometry is $(\chi, \xi) = (75^\circ,5^\circ)$. See text for all details.}
    \label{fig:mock_spectra_0720}
\end{figure*}

Figure \ref{fig:mock_spectra_0720} shows the expected counts in each detector computed 
assuming emission from either an atmosphere or a condensate. The viewing geometry is $(\chi, \xi) = (75^\circ,5^\circ)$, as inferred by \cite{haberl_evidence_2006}. 
In the mock spectra corresponding to the atmospheric model with no mode conversion,
we can see a clear dip in the counts of detectors 2 and 3 at $\sim 50$ Å in both the QED on and QED off cases. 
This is due to the  expected proton cyclotron feature at $\sim 270 \ \mathrm{eV}$ associated to the polar cap with the weakest magnetic field.  Instead, from the detector counts alone, for all of the condensed surface emission cases and for the atmospheric model with complete mode conversion, 
there is no clear indication of any feature in the detector energy band.

\begin{figure*}
    \centering
    \includegraphics[width=1\linewidth]{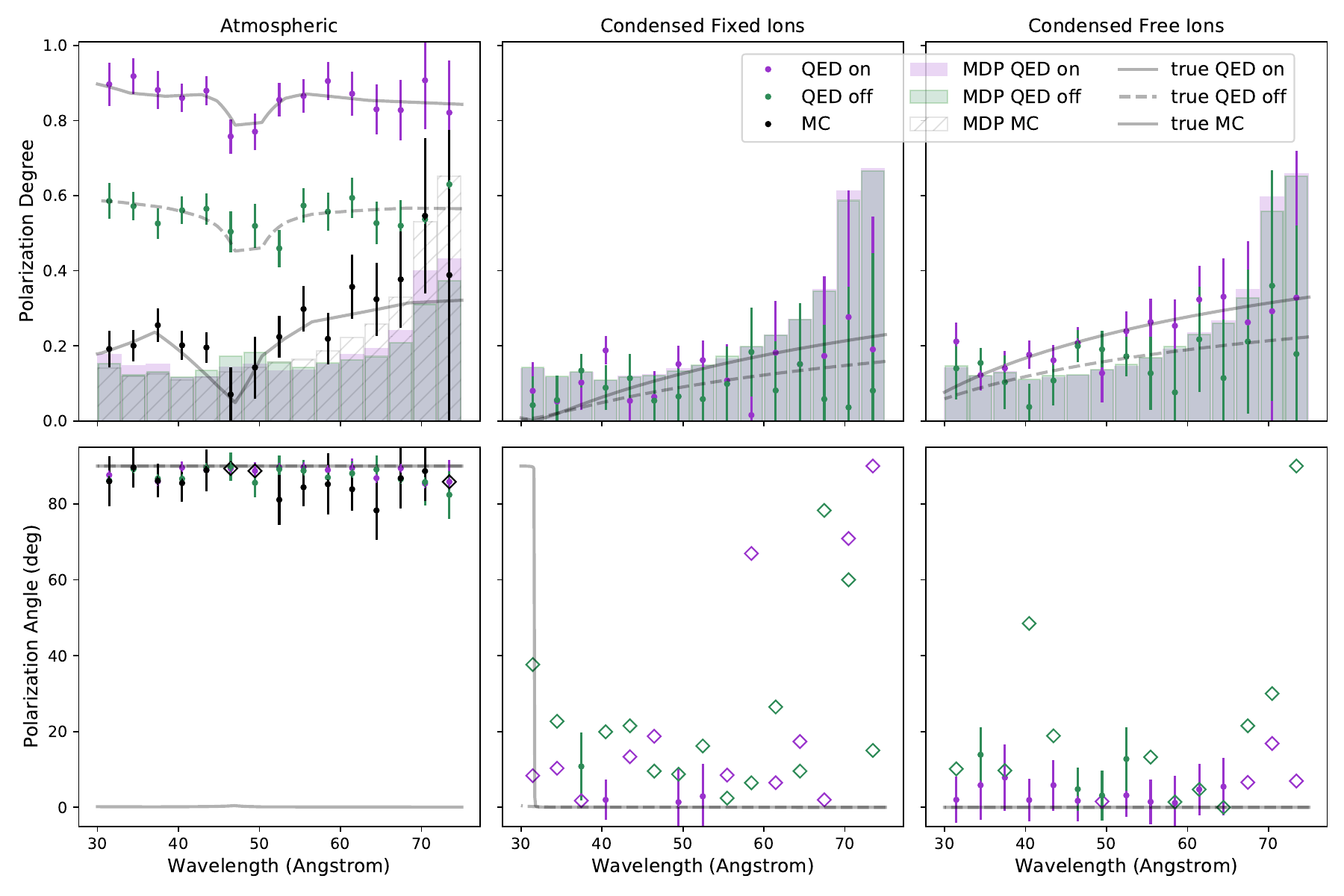}
    \caption{Expected polarization degree and angle computed assuming 
    the models as in figure \ref{fig:mock_spectra_0720}, for RX~J0720.
    The points show the simulated data with $1 \sigma$-errors, the solid lines the original models, and the shaded areas the MDP for each mock observation. 
    When the polarization degree is below MDP, the error bars are shown at $90\%$ c.l., while  the diamonds represent an unconstrained polarization angle.}
    \label{fig:0720_pol}
\end{figure*}

The expected polarization properties, computed by combining the three detectors, are shown in Figure \ref{fig:0720_pol}. We start by discussing the expectations of atmospheric models.  The model computed with no mode conversion, 
once again, produces very high polarization across the entire detector energy range in the QED on case, with a frequency integrated $\mathrm{PD} = (86.5 \pm 1.3)\,\%$. In this scenario, a clear dip in $\mathrm{PD}$  at the proton cyclotron energy ($\sim 50$ Å) is visible \footnote{In the case of pure plasma (no vacuum corrections) with QED effects accounted for in the photon propagation toward the observer, the proton cyclotron dip is even more prominent, with the $\mathrm{PD}$ dipping to $\sim70\%$ at the cyclotron energy.}. This dip is less evident in the corresponding QED off model, in which the polarization degree remains seemingly constant with energy around a frequency-integrated value of $\mathrm{PD} = 55.2 \pm 1.2\,\%$ (for this viewing geometry). 

We also tested models with partial mode conversion, varying the assumption on $P_\mathrm{con}^\mathrm{th}$, and found qualitatively similar results to the QED on case without mode conversion, with slightly lower values for the degree of polarization.

The complete mode conversion scenario results in a relatively low polarization,  $\mathrm{PD} = (24.9 \pm 1.3)\, \%$ integrated over the instrument energy band, significantly reduced compared to the no mode conversion scenarios. The proton cyclotron absorption feature is clearly seen in the spectrum, although the drop in polarization causes the $\mathrm{PD}$ to fall below the MDP at this wavelength. However, for most of the wavelength range (excluding the highest wavelengths), the expected polarization remains above the MDP, being measurable by REDSoX.

In all these three cases, the polarization angle remains constant across the whole band, around $\sim 90^\circ$, indicating that, at least for this choice of thermal map and geometry, we do not expect to detect the switch in dominant polarization mode due to mode conversion in the detector band. 

The $90^\circ$ swing in $\mathrm{PA}$ due to mode conversion is not visible because, according to our simulations, at $2\times10^{14} \, \mathrm{G}$, the magnetic field of the small patch is slightly too high, so the swing occurs below $0.2\, \mathrm{keV}$. The emission is instead dominated by the radiation emitted by the rest of the polar cap. Interestingly, the emission from the polar cap and that from the strongly magnetized patch are dominated by opposite polarization modes (O and X respectively) in the REDSoX band, therefore the inclusion of the patch in the simulation  leads to a slightly depolarized signal. 

In principle, one might expect the detectability of a $90^\circ$ swing in $\mathrm{PA}$ in phase resolved observations. However, we found that the larger contribution from the region with the lower field always  dominates and  washes out the contribution from the magnetic loop. Only when it is assumed that the loop covers a large region (almost half) of the polar cap,  then a swing is evident in the phase-resolved polarization spectrum. Noticeably, the origin of this swing is different in models with different magnetic fields strengths in the small patch: if $B \geq  10^{14} \, \mathrm{G}$ it is due to the  fact that two different modes dominates in the different emitting regions, while when  the strength of the field in the magnetic loop is reduced to $\lesssim 10^{14} \, \mathrm{G}$, the  $90^\circ$ swing in $\mathrm{PA}$ due to QED mode conversion becomes visible in phase resolved data.

On the other hand, models based on condensed surface emission produce significantly lower polarization in all cases and have no visible proton cyclotron feature in the spectra. In the limiting case of fixed ions, both QED on and QED off, the polarization degree remains below MDP for almost all of the wavelength bins, and the polarization angle is therefore unconstrained. However, the energy-integrated  polarization, $\mathrm{PD} = (10.1 \pm 1.4)\,\%$ and $( 7.3 \pm 1.2)\,\%$, for QED on and off, respectively, is above the energy-integrated MDP$=4.9\%$. For the free ion case, only in the QED off case the expected $\mathrm{PD}$ is below the MPD in most wavelength bins; in this case the  energy-averaged is $\mathrm{PD} = (12.0 \pm 1.3)\,\%$. When QED effects are taken into account, the case of free ions shows a clear increase in polarization with increasing wavelength, from $\sim 10\%$ to $\sim 30\%$. In this case, the frequency average $\mathrm{PD} = (19.0 \pm 1.3)\,\%$.

\section{Discussion and Conclusions} \label{sec:discussion}

The observation of X-ray polarization below $1 \, \mathrm{keV}$ is a crucial path to advancing our understanding of XDINS and magnetar phenomena. A space mission hosting an instrument with the same or similar design as the REDSoX mission, observing in the $0.2$--$0.4\, \mathrm{keV}$ range, such as the proposed global orbiting mission, GOSoX, will allow us to study the magnetic fields and external layers of isolated neutron stars, identify the emission mechanisms, and, significantly, may even test QED vacuum birefringence. Polarimetric data in the soft X-ray band are therefore key, in themselves and also when used in combination with those from other polarimetric instruments working in different wavebands, to guiding our knowledge of the physics of these extreme sources.

Under standard assumptions of a pure plasma, fully ionized H, passive cooler  magnetized atmosphere, thermal emission is well known to be highly polarized, at $\gtrsim 80\%$,  and dominated by X mode photons. 
However, mode conversion at the QED vacuum resonance can greatly affect the polarization signal. We found that, for dipolar magnetic field strengths $B\sim 10^{14}\, \mathrm{G}$, complete mode conversion results in the presence of a switch in dominant polarization mode (implying a $90^\circ$ swing in polarization angle) that falls in the $0.2$--$0.4\, \mathrm{keV}$ energy range. This rotation in polarization angle can be resolved by the REDSoX instrument with $\sim25,000$ counts. An observation of this kind from a source would therefore provide a robust observational test for the presence of QED vacuum birefringence, as well as put constraints on the physics of mode conversion and on the magnetic field strength of the source.  
Additionally, the non-detection of the angle swing would not rule out QED: in fact models computed assuming partial mode conversion (for the same magnetic field strength) only result, in the REDSoX range, in a reduced degree of polarization with respect to those computed turning off QED effects completely in the atmospheric model (but keeping them in the photon transport from the source to the observer). However, in all these cases, the degree of polarization is still expected to be well above that expected in the case of condensed surface emission (which is $\lesssim 20\%$). 

Proton cyclotron absorption features in the polarization spectrum can also originate in the star atmosphere, and they are expected to fall in the $0.2$--$0.4\, \mathrm{keV}$ energy range 
for $B\sim 5\times10^{13} \, \mathrm{G}$. 
We found that observations from an instrument with the same design as REDSoX can resolve these absorption features (or, in general, spectral absorption features of similar strength) with high statistical significance when collecting $\gtrsim 25,000$ counts from a source.

In addition to the specific spectral features discussed above, the instrument will be able to detect smooth changes in the polarization signal with energy/wavelength (e.g. increases/decreases in PD across the band), which will further inform the interpretation of the source emission.
In all models considered in our simulations, the signal emerging from a condensed surface exhibits a  significantly lower polarization degree ($\mathrm{PD} \lesssim 20 \%$) compared to that caused by atmospheric emission, showing that the design of the REDSoX instrument will allow us to identify the emission mechanisms and compositions of the external surface layer of the neutron star.

We also considered two case studies, focusing on two XDINS targets and producing simulations based on some examples of very simplified thermal maps, assumed in a way to mimic, at least in the first approximation, the X-ray spectral properties of the sources. 
Quite interestingly, in the case of RX~J1856, we found that all the atmospheric models we considered (independently on the treatment of mode conversion at the vacuum resonance) concur in predicting a very high degree of polarization in the observed REDSoX emission ($\mathrm{PD} \sim 95\%$) if, and only if, QED vacuum birefringence is at work in the vacuum around the star. An observation of such high polarization from this source will therefore allow for a long-awaited observational evidence of the existence of QED vacuum birefringence.

In the case of the second target, RX~J0720, an absorption feature at $\sim 270 \ \mathrm{eV}$ has previously been observed in the spectrum, and this falls in the REDSoX energy range. Our models based on atmospheric emission exhibit this feature in the polarization spectrum, as caused by the proton cyclotron resonance, and it is particularly prominent for models computed including QED effects. We also attempted to model the presence of highly magnetized loops near the star surface, as inferred for this target. We found that 
the loop must be relatively large to be observable (in our simulation covering a region as large as half of the cap that extends up to $\sim 30^\circ$ in colatitude). However, would this be the case, then for some particular emission models,  its presence  may be detectable in the REDSoX band through phase dependent observations of a $90^\circ$ swing in PA. Although the nature of the swing depends on the magnetic field strength, its presence is always caused, directly or indirectly, by the occurrence of QED complete mode conversion at the vacuum resonance. 

The geometry and magnetic field strength of an isolated neutron star is typically not well constrained. A comprehensive investigation into the affects on the resulting uncertainties by varying the pulsar geometry and magnetic field strength would be informative, but would warrant a systematic analysis and is beyond the scope of this work. Here, we aim to highlight what can be detected with an instrument such as REDSoX. To that end, for the two example cases in section \ref{sec:application}, the models, viewing angles and magnetic field strengths were chosen to be most favorable for detecting the respective features. For the two case studies we used geometries and field strengths previously quoted in the literature. For our analysis of J1856, we rely on the $\chi$ and $\xi$ values quoted in \cite{mignani_evidence_2017} where the pulsed fraction and polarization properties of the source are considered together. In this case the uncertainties in measurements of $\chi$ and $\xi$ are relatively small, and the differences in polarization only varies modestly. In contrast, in the case of J0720, the viewing geometry is much less constrained. Here, we have used the $\chi$, $\xi$ values from \cite{haberl_evidence_2006} which are favorable for our investigation.

Phase-resolved polarimetry is important for constraining the geometry of the surface emission, in particular because if the behavior of the polarization angle is not well-reproduced by the rotating vector model (RVM) (and the magnetic field is dipolar) it may imply the presence of a non-axisymmetric thermal map \cite[see supplementary material from][]{taverna_x-ray_2020}. For the models and viewing angles used in this work, a phase resolved analysis shows only very small variations in polarization degree with phase, while polarization angles follow a smooth sinusoidal oscillation, in agreement with the RVM. 

In principle, modeling observed data with the RVM has been often proposed as an indirect test of QED vacuum birefringence. However, this test is not meaningful in case of sources for which the magnetic field is intrinsically close to a dipole, such as the ones considered in this paper. 
For many cases, as highlighted above in the case of J1856, phase-resolved observations will be crucial for further constraining the geometries involved, and our understanding of isolated neutron star sources more generally. A longer, orbital mission, equipped with an instrument similar to REDSoX would undoubtedly benefit from high time resolution, allowing us to make the necessary measurements to constrain emission and viewing geometries, as well as to resolve phase dependent absorption features (such as that observed in the magnetar IE 2259+586 \citep{heyl_detection_2024}).

We caveat that our atmospheric models assume a fully ionized atmosphere, and, consequently, exhibit cyclotron lines as the only absorption features in the X-ray spectrum. However, this is a simplification and the coldest atmospheric layers may be expected to be partially ionized. The inclusion of atomic transitions and resonances complicates the calculation of radiative transfer and may affect the polarization of the emitted radiation \cite[see e.g.][for details]{potekhin_electromagnetic_2004, potekhin_atmospheres_2014}. The inclusion of partial ionization to our atmosphere models is a matter of ongoing work. 

Although based on simple assumptions, our simulations of a few illustrative cases show that a fully-fledged  orbital mission utilizing the REDSoX instrument design will undoubtedly lend itself to important scientific advancement, specifically in the studies of neutron stars and highly magnetized environments.

\begin{acknowledgments}
RMEK would like to thank the MIT Kavli Institute for supporting and hosting her visit and The Science and Technology Facilities Council (STFC) for financial support via a Long Term Attachment (LTA) grant and a PhD studentship (grant number ST/W507891/1). This work was supported in part by MIT Corporate Relations and its Industrial Liaison Program (ILP). HLM and SR were supported in part by NASA grant 80NSSC23K0644 for development of the REDSoX Polarimeter. NB is supported by STFC grant ST/Y001060/1. RTa and RTu are partially supported by the PRIN grant 2022LWPEXW of the Italian Ministry of University and Research (MUR). DGC acknowledges support from the Agence Nationale de la Recherche (ANR) under grant number ANR-20-CE31-0010 (MORPHER). We would like to thank the anonymous referee for their constructive and helpful feedback.
\end{acknowledgments}

%

\vspace{5mm}


\software{tbabs, https://pulsar.sternwarte.uni-erlangen.de/wilms/research/tbabs/ \citep{wilms_absorption_2000}
Other supplementary material or numerical routines can be provided upon request.
          }





\bibliography{references}{}

\begin{thebibliography}{}
\expandafter\ifx\csname natexlab\endcsname\relax\def\natexlab#1{#1}\fi
\providecommand{\url}[1]{\href{#1}{#1}}
\providecommand{\dodoi}[1]{doi:~\href{http://doi.org/#1}{\nolinkurl{#1}}}
\providecommand{\doeprint}[1]{\href{http://ascl.net/#1}{\nolinkurl{http://ascl.net/#1}}}
\providecommand{\doarXiv}[1]{\href{https://arxiv.org/abs/#1}{\nolinkurl{https://arxiv.org/abs/#1}}}

\bibitem[{Adler(1971)}]{adler_photon_1971}
Adler, S.~L. 1971, Annals of Physics, 67, 599, \dodoi{10.1016/0003-4916(71)90154-0}

\bibitem[{Blandford {et~al.}(2019)Blandford, Meier, \& Readhead}]{blandford_relativistic_2019}
Blandford, R., Meier, D., \& Readhead, A. 2019, Annual Review of Astronomy and Astrophysics, 57, 467, \dodoi{10.1146/annurev-astro-081817-051948}

\bibitem[{Borghese {et~al.}(2015)Borghese, Rea, Zelati, Tiengo, \& Turolla}]{borghese_discovery_2015}
Borghese, A., Rea, N., Zelati, F.~C., Tiengo, A., \& Turolla, R. 2015, The Astrophysical Journal, 807, L20, \dodoi{10.1088/2041-8205/807/1/L20}

\bibitem[{Brinkmann(1980)}]{brinkmann_thermal_1980}
Brinkmann, W. 1980, Astronomy and Astrophysics, 82, 352.
\newblock \url{https://ui.adsabs.harvard.edu/abs/1980A&A....82..352B}

\bibitem[{Chauvin {et~al.}(2017)Chauvin, Friis, Jackson, Kawano, Kiss, Mikhalev, Ohashi, Stana, Takahashi, \& Pearce}]{chauvin_calibration_2017}
Chauvin, M., Friis, M., Jackson, M., {et~al.} 2017, Nuclear Instruments and Methods in Physics Research Section A: Accelerators, Spectrometers, Detectors and Associated Equipment, 859, 125, \dodoi{10.1016/j.nima.2017.03.027}

\bibitem[{De~Grandis {et~al.}(2021)De~Grandis, Taverna, Turolla, Gnarini, Popov, Zane, \& Wood}]{de_grandis_x-ray_2021}
De~Grandis, D., Taverna, R., Turolla, R., {et~al.} 2021, The Astrophysical Journal, 914, 118, \dodoi{10.3847/1538-4357/abfdac}

\bibitem[{De Grandis {et~al.}(2022)De Grandis, Rigoselli, Mereghetti, Younes, Pizzochero, Taverna, Tiengo, Turolla, \& Zane}]{degrandis_two_2022}
De Grandis, D., Rigoselli, M., Mereghetti, S., {et~al.} 2022, Monthly Notices of the Royal Astronomical Society, 516, 4932, \dodoi{10.1093/mnras/stac2587}

\bibitem[{Feng {et~al.}(2019)Feng, Jiang, Minuti, Wu, Jung, Yang, Citraro, Nasimi, Yu, Jin, Huang, Zeng, An, Baldini, Bellazzini, Brez, Latronico, Sgrò, Spandre, Pinchera, Muleri, Soffitta, \& Costa}]{feng_polarlight_2019}
Feng, H., Jiang, W., Minuti, M., {et~al.} 2019, Experimental Astronomy, 47, 225, \dodoi{10.1007/s10686-019-09625-z}

\bibitem[{Garner {et~al.}(2024)Garner, Marshall, Trowbridge~Heine, McNeil, Schulz, Heilmann, Günther, Juneau, LaMarr, Metivier, Ravi, Kothnur, Bongiorno, \& Gullikson}]{garner_current_2024}
Garner, A., Marshall, H.~L., Trowbridge~Heine, S.~N., {et~al.} 2024, in Space {Telescopes} and {Instrumentation} 2024: {Ultraviolet} to {Gamma} {Ray}, ed. J.-W.~A. Den~Herder, K.~Nakazawa, \& S.~Nikzad (Yokohama, Japan: SPIE), 277, \dodoi{10.1117/12.3020172}

\bibitem[{Gnedin \& Pavlov(1974)}]{gnedin_transfer_1974}
Gnedin, Y.~N., \& Pavlov, G.~G. 1974, Soviet Journal of Experimental and Theoretical Physics, 38, 903.
\newblock \url{https://ui.adsabs.harvard.edu/abs/1974JETP...38..903G}

\bibitem[{González~Caniulef {et~al.}(2016)González~Caniulef, Zane, Taverna, Turolla, \& Wu}]{gonzalez_caniulef_polarized_2016}
González~Caniulef, D., Zane, S., Taverna, R., Turolla, R., \& Wu, K. 2016, Monthly Notices of the Royal Astronomical Society, 459, 3585, \dodoi{10.1093/mnras/stw804}

\bibitem[{González-Caniulef {et~al.}(2019)González-Caniulef, Zane, Turolla, \& Wu}]{gonzalez-caniulef_atmosphere_2019}
González-Caniulef, D., Zane, S., Turolla, R., \& Wu, K. 2019, Monthly Notices of the Royal Astronomical Society, 483, 599, \dodoi{10.1093/mnras/sty3159}

\bibitem[{Greenstein \& Hartke(1983)}]{greenstein_pulselike_1983}
Greenstein, G., \& Hartke, G.~J. 1983, The Astrophysical Journal, 271, 283, \dodoi{10.1086/161195}

\bibitem[{Haberl {et~al.}(2006)Haberl, Turolla, De~Vries, Zane, Vink, Méndez, \& Verbunt}]{haberl_evidence_2006}
Haberl, F., Turolla, R., De~Vries, C.~P., {et~al.} 2006, Astronomy \& Astrophysics, 451, L17, \dodoi{10.1051/0004-6361:20065093}

\bibitem[{Haberl {et~al.}(2004)Haberl, Zavlin, Trümper, \& Burwitz}]{haberl_phase-dependent_2004}
Haberl, F., Zavlin, V.~E., Trümper, J., \& Burwitz, V. 2004, Astronomy \& Astrophysics, 419, 1077, \dodoi{10.1051/0004-6361:20034129}

\bibitem[{Heine {et~al.}(2024)Heine, Marshall, Schneider, Lamar, Kothnur, \& Garner}]{heine_characterization_2024}
Heine, S. N.~T., Marshall, H.~L., Schneider, B., {et~al.} 2024, Characterization of {X}-ray {Detectors} in the {MIT} {X}-ray {Polarimetry} {Beamline},  arXiv, \dodoi{10.48550/ARXIV.2408.11168}

\bibitem[{Heisenberg \& Euler(1936)}]{heisenberg_folgerungen_1936}
Heisenberg, W., \& Euler, H. 1936, Zeitschrift for Physik, 98, 714, \dodoi{10.1007/BF01343663}

\bibitem[{Heyl {et~al.}(2024)Heyl, Taverna, Turolla, Israel, Ng, Kirmizibayrak, González-Caniulef, Caiazzo, Zane, Ehlert, Negro, Agudo, Antonelli, Bachetti, Baldini, Baumgartner, Bellazzini, Bianchi, Bongiorno, Bonino, Brez, Bucciantini, Capitanio, Castellano, Cavazzuti, Chen, Ciprini, Costa, De~Rosa, Del~Monte, Di~Gesu, Di~Lalla, Di~Marco, Donnarumma, Doroshenko, Dovčiak, Enoto, Evangelista, Fabiani, Ferrazzoli, Garcia, Gunji, Hayashida, Iwakiri, Jorstad, Kaaret, Karas, Kislat, Kitaguchi, Kolodziejczak, Krawczynski, La~Monaca, Latronico, Liodakis, Maldera, Manfreda, Marin, Marinucci, Marscher, Marshall, Massaro, Matt, Mitsuishi, Mizuno, Muleri, Ng, O'Dell, Omodei, Oppedisano, Papitto, Pavlov, Peirson, Perri, Pesce-Rollins, Petrucci, Pilia, Possenti, Poutanen, Puccetti, Ramsey, Rankin, Ratheesh, Roberts, Romani, Sgrò, Slane, Soffitta, Spandre, Swartz, Tamagawa, Tavecchio, Tawara, Tennant, Thomas, Tombesi, Trois, Tsygankov, Vink, Weisskopf, Wu, \& Xie}]{heyl_detection_2024}
Heyl, J., Taverna, R., Turolla, R., {et~al.} 2024, Monthly Notices of the Royal Astronomical Society, 527, 12219, \dodoi{10.1093/mnras/stad3680}

\bibitem[{Ho \& Lai(2003)}]{ho_ii_2003}
Ho, W. C.~G., \& Lai, D. 2003, Monthly Notices of the Royal Astronomical Society, 338, 233, \dodoi{10.1046/j.1365-8711.2003.06047.x}

\bibitem[{Ho {et~al.}(2003)Ho, Lai, Potekhin, \& Chabrier}]{ho_iii_2003}
Ho, W. C.~G., Lai, D., Potekhin, A.~Y., \& Chabrier, G. 2003, The Astrophysical Journal, 599, 1293, \dodoi{10.1086/379507}

\bibitem[{ISRO(2024)}]{isro_xposat_2024}
ISRO. 2024, {XPoSat}.
\newblock \url{https://www.isro.gov.in/XPoSat.html}

\bibitem[{Kelly {et~al.}(2024{\natexlab{a}})Kelly, González-Caniulef, Zane, Turolla, \& Taverna}]{kelly_x-ray_2024}
Kelly, R. M.~E., González-Caniulef, D., Zane, S., Turolla, R., \& Taverna, R. 2024{\natexlab{a}}, Monthly Notices of the Royal Astronomical Society, 534, 1355, \dodoi{10.1093/mnras/stae2163}

\bibitem[{Kelly {et~al.}(2024{\natexlab{b}})Kelly, Zane, Turolla, \& Taverna}]{kelly_x-ray_2024-1}
Kelly, R. M.~E., Zane, S., Turolla, R., \& Taverna, R. 2024{\natexlab{b}}, Monthly Notices of the Royal Astronomical Society, 528, 3927, \dodoi{10.1093/mnras/stae159}

\bibitem[{Lai(2023)}]{lai_ixpe_2023}
Lai, D. 2023, Proceedings of the National Academy of Sciences, 120, e2216534120, \dodoi{10.1073/pnas.2216534120}

\bibitem[{Lai \& Ho(2003)}]{lai_transfer_2003}
Lai, D., \& Ho, W. C.~G. 2003, The Astrophysical Journal, 588, 962, \dodoi{10.1086/374334}

\bibitem[{Lai \& Salpeter(1997)}]{lai_hydrogen_1997}
Lai, D., \& Salpeter, E.~E. 1997, The Astrophysical Journal, 491, 270, \dodoi{10.1086/304937}

\bibitem[{Lloyd(2003)}]{lloyd_model_2003}
Lloyd, D.~A. 2003, Model atmospheres and thermal spectra of magnetized neutron stars, \dodoi{10.48550/ARXIV.ASTRO-PH/0303561}

\bibitem[{Marshall {et~al.}(2024)Marshall, Heine, Garner, Masterson, Guenther, Heilmann, Bongiorno, \& Gullikson}]{marshall_rocket_2024}
Marshall, H., Heine, S., Garner, A., {et~al.} 2024, Multifrequency Behaviour of High Energy Cosmic Sources XIV, 76.
\newblock \url{https://ui.adsabs.harvard.edu/abs/2024mbhe.confE..76M/abstract}

\bibitem[{Marshall {et~al.}(2023)Marshall, Heine, Garner, Masterson, \& Heilmann}]{marshall_status_2023}
Marshall, H.~L., Heine, S., Garner, A., Masterson, R., \& Heilmann, R. 2023, in {UV}, {X}-{Ray}, and {Gamma}-{Ray} {Space} {Instrumentation} for {Astronomy} {XXIII}, ed. O.~H. Siegmund \& K.~Hoadley (San Diego, United States: SPIE), 40, \dodoi{10.1117/12.2677873}

\bibitem[{Marshall {et~al.}(2018)Marshall, Günther, Heilmann, Schulz, Egan, Hellickson, Heine, Windt, Gullikson, \& Ramsey}]{marshall_design_2018}
Marshall, H.~L., Günther, H.~M., Heilmann, R.~K., {et~al.} 2018, Journal of Astronomical Telescopes, Instruments, and Systems, 4, 1, \dodoi{10.1117/1.JATIS.4.1.011005}

\bibitem[{Marshall {et~al.}(2021)Marshall, Heine, Davidson, Garner, Gullikson, Günther, Leitz, Masterson, Miller, Stenzel, Zhang, Boissay-Malaquin, Caiazzo, Chakrabarty, Gallo, Heilmann, Heyl, Kara, \& Schulz}]{marshall_globe_2021}
Marshall, H.~L., Heine, S., Davidson, R., {et~al.} 2021, in Optics for {EUV}, {X}-{Ray}, and {Gamma}-{Ray} {Astronomy} {X}, ed. G.~Pareschi, S.~L. O'Dell, \& J.~A. Gaskin (San Diego, United States: SPIE), 67, \dodoi{10.1117/12.2596186}

\bibitem[{Mignani {et~al.}(2017)Mignani, Testa, González~Caniulef, Taverna, Turolla, Zane, \& Wu}]{mignani_evidence_2017}
Mignani, R.~P., Testa, V., González~Caniulef, D., {et~al.} 2017, Monthly Notices of the Royal Astronomical Society, 465, 492, \dodoi{10.1093/mnras/stw2798}

\bibitem[{Novick {et~al.}(1972)Novick, Weisskopf, Berthelsdorf, Linke, \& Wolff}]{novick_detection_1972}
Novick, R., Weisskopf, M.~C., Berthelsdorf, R., Linke, R., \& Wolff, R.~S. 1972, The Astrophysical Journal, 174, L1, \dodoi{10.1086/180938}

\bibitem[{Page \& Sarmiento(1996)}]{page_surface_1996}
Page, D., \& Sarmiento, A. 1996, The Astrophysical Journal, 473, 1067, \dodoi{10.1086/178216}

\bibitem[{Pizzocaro {et~al.}(2019)Pizzocaro, Tiengo, Mereghetti, Turolla, Esposito, Stella, Zane, Rea, Coti~Zelati, \& Israel}]{pizzocaro_detailed_2019}
Pizzocaro, D., Tiengo, A., Mereghetti, S., {et~al.} 2019, Astronomy \& Astrophysics, 626, A39, \dodoi{10.1051/0004-6361/201834784}

\bibitem[{Popov {et~al.}(2017)Popov, Taverna, \& Turolla}]{popov_probing_2017}
Popov, S.~B., Taverna, R., \& Turolla, R. 2017, Monthly Notices of the Royal Astronomical Society, 464, 4390, \dodoi{10.1093/mnras/stw2681}

\bibitem[{Potekhin(2014)}]{potekhin_atmospheres_2014}
Potekhin, A.~Y. 2014, Physics-Uspekhi, 57, 735, \dodoi{10.3367/UFNe.0184.201408a.0793}

\bibitem[{Potekhin {et~al.}(2004)Potekhin, Lai, Chabrier, \& Ho}]{potekhin_electromagnetic_2004}
Potekhin, A.~Y., Lai, D., Chabrier, G., \& Ho, W. C.~G. 2004, The Astrophysical Journal, 612, 1034, \dodoi{10.1086/422679}

\bibitem[{Potekhin {et~al.}(2012)Potekhin, Suleimanov, Van~Adelsberg, \& Werner}]{potekhin_radiative_2012}
Potekhin, A.~Y., Suleimanov, V.~F., Van~Adelsberg, M., \& Werner, K. 2012, Astronomy \& Astrophysics, 546, A121, \dodoi{10.1051/0004-6361/201219747}

\bibitem[{Rigoselli {et~al.}(2024)Rigoselli, Taverna, Mereghetti, Turolla, Israel, Zane, Marra, Muleri, Borghese, Zelati, De~Grandis, Imbrogno, Kelly, Esposito, \& Rea}]{rigoselli_ixpe_2024}
Rigoselli, M., Taverna, R., Mereghetti, S., {et~al.} 2024, {IXPE} detection of highly polarized {X}-rays from the magnetar {1E} 1841-045,  arXiv, \dodoi{10.48550/ARXIV.2412.15811}

\bibitem[{Sartore {et~al.}(2012)Sartore, Tiengo, Mereghetti, De~Luca, Turolla, \& Haberl}]{sartore_spectral_2012}
Sartore, N., Tiengo, A., Mereghetti, S., {et~al.} 2012, Astronomy \& Astrophysics, 541, A66, \dodoi{10.1051/0004-6361/201118489}

\bibitem[{Smith \& Courtier(1976)}]{smith_ariel_1976}
Smith, J.~F., \& Courtier, G.~M. 1976, Proceedings of the Royal Society of London. A. Mathematical and Physical Sciences, 350, 421, \dodoi{10.1098/rspa.1976.0115}

\bibitem[{Stewart {et~al.}(2024)Stewart, Younes, Harding, Wadiasingh, Baring, Negro, Strohmayer, Ho, Ng, Arzoumanian, Thi, Di~Lalla, Enoto, Gendreau, Hu, van Kooten, Kouveliotou, \& McEwen}]{stewart_x-ray_2024}
Stewart, R., Younes, G., Harding, A., {et~al.} 2024, X-ray polarization of the magnetar {1E} 1841-045 in outburst,  arXiv, \dodoi{10.48550/ARXIV.2412.16036}

\bibitem[{Taverna {et~al.}(2015)Taverna, Turolla, Gonzalez~Caniulef, Zane, Muleri, \& Soffitta}]{taverna_polarization_2015}
Taverna, R., Turolla, R., Gonzalez~Caniulef, D., {et~al.} 2015, Monthly Notices of the Royal Astronomical Society, 454, 3254, \dodoi{10.1093/mnras/stv2168}

\bibitem[{Taverna {et~al.}(2020)Taverna, Turolla, Suleimanov, Potekhin, \& Zane}]{taverna_x-ray_2020}
Taverna, R., Turolla, R., Suleimanov, V., Potekhin, A.~Y., \& Zane, S. 2020, Monthly Notices of the Royal Astronomical Society, 492, 5057, \dodoi{10.1093/mnras/staa204}

\bibitem[{Taverna {et~al.}(2022)Taverna, Turolla, Muleri, Heyl, Zane, Baldini, González-Caniulef, Bachetti, Rankin, Caiazzo, Di~Lalla, Doroshenko, Errando, Gau, Kırmızıbayrak, Krawczynski, Negro, Ng, Omodei, Possenti, Tamagawa, Uchiyama, Weisskopf, Agudo, Antonelli, Baumgartner, Bellazzini, Bianchi, Bongiorno, Bonino, Brez, Bucciantini, Capitanio, Castellano, Cavazzuti, Ciprini, Costa, De~Rosa, Del~Monte, Di~Gesu, Di~Marco, Donnarumma, Dovčiak, Ehlert, Enoto, Evangelista, Fabiani, Ferrazzoli, Garcia, Gunji, Hayashida, Iwakiri, Jorstad, Karas, Kitaguchi, Kolodziejczak, La~Monaca, Latronico, Liodakis, Maldera, Manfreda, Marin, Marinucci, Marscher, Marshall, Matt, Mitsuishi, Mizuno, Ng, O’Dell, Oppedisano, Papitto, Pavlov, Peirson, Perri, Pesce-Rollins, Pilia, Poutanen, Puccetti, Ramsey, Ratheesh, Romani, Sgrò, Slane, Soffitta, Spandre, Tavecchio, Tawara, Tennant, Thomas, Tombesi, Trois, Tsygankov, Vink, Wu, \& Xie}]{taverna_polarized_2022}
Taverna, R., Turolla, R., Muleri, F., {et~al.} 2022, Science, 378, 646, \dodoi{10.1126/science.add0080}

\bibitem[{Tiengo \& Mereghetti(2007)}]{tiengo_xmm-newton_2007}
Tiengo, A., \& Mereghetti, S. 2007, The Astrophysical Journal, 657, L101, \dodoi{10.1086/513143}

\bibitem[{Turolla(2009)}]{turolla_isolated_2009}
Turolla, R. 2009, Astrophysics and {Space} {Science} {Library}, Vol. 357, Isolated {Neutron} {Stars}: {The} {Challenge} of {Simplicity} (Springer Berlin Heidelberg)

\bibitem[{Turolla {et~al.}(2015)Turolla, Zane, \& Watts}]{turolla_magnetars_2015}
Turolla, R., Zane, S., \& Watts, A.~L. 2015, Reports on Progress in Physics, 78, 116901, \dodoi{10.1088/0034-4885/78/11/116901}

\bibitem[{Turolla {et~al.}(2023)Turolla, Taverna, Israel, Muleri, Zane, Bachetti, Heyl, Marco, Gau, Krawczynski, Ng, Possenti, Poutanen, Baldini, Matt, Negro, Agudo, Antonelli, Baumgartner, Bellazzini, Bianchi, Bongiorno, Bonino, Brez, Bucciantini, Capitanio, Castellano, Cavazzuti, Chen, Ciprini, Costa, Rosa, Monte, Gesu, Lalla, Donnarumma, Doroshenko, Dovčiak, Ehlert, Enoto, Evangelista, Fabiani, Ferrazzoli, Garcia, Gunji, Hayashida, Iwakiri, Jorstad, Kaaret, Karas, Kislat, Kitaguchi, Kolodziejczak, Monaca, Latronico, Liodakis, Maldera, Manfreda, Marin, Marinucci, Marscher, Marshall, Massaro, Mitsuishi, Mizuno, Ng, O’Dell, Omodei, Oppedisano, Papitto, Pavlov, Peirson, Perri, Pesce-Rollins, Petrucci, Pilia, Puccetti, Ramsey, Rankin, Ratheesh, Roberts, Romani, Sgró, Slane, Soffitta, Spandre, Swartz, Tamagawa, Tavecchio, Tawara, Tennant, Thomas, Tombesi, Trois, Tsygankov, Vink, Weisskopf, Wu, \& Xie}]{turolla_ixpe_2023}
Turolla, R., Taverna, R., Israel, G.~L., {et~al.} 2023, The Astrophysical Journal, 954, 88, \dodoi{10.3847/1538-4357/aced05}

\bibitem[{Van~Adelsberg {et~al.}(2005)Van~Adelsberg, Lai, Potekhin, \& Arras}]{van_adelsberg_radiation_2005}
Van~Adelsberg, M., Lai, D., Potekhin, A.~Y., \& Arras, P. 2005, The Astrophysical Journal, 628, 902, \dodoi{10.1086/430871}

\bibitem[{Van~Kerkwijk \& Kaplan(2007)}]{van_kerkwijk_isolated_2007}
Van~Kerkwijk, M.~H., \& Kaplan, D.~L. 2007, Astrophysics and Space Science, 308, 191, \dodoi{10.1007/s10509-007-9343-9}

\bibitem[{Verner {et~al.}(1996)Verner, Ferland, Korista, \& Yakovlev}]{verner_atomic_1996}
Verner, D.~A., Ferland, G.~J., Korista, K.~T., \& Yakovlev, D.~G. 1996, The Astrophysical Journal, 465, 487, \dodoi{10.1086/177435}

\bibitem[{Viganò {et~al.}(2014)Viganò, Perna, Rea, \& Pons}]{vigano_spectral_2014}
Viganò, D., Perna, R., Rea, N., \& Pons, J.~A. 2014, Monthly Notices of the Royal Astronomical Society, 443, 31, \dodoi{10.1093/mnras/stu1109}

\bibitem[{Walter {et~al.}(2010)Walter, Eisenbeiß, Lattimer, Kim, Hambaryan, \& Neuhäuser}]{walter_revisiting_2010}
Walter, F.~M., Eisenbeiß, T., Lattimer, J.~M., {et~al.} 2010, The Astrophysical Journal, 724, 669, \dodoi{10.1088/0004-637X/724/1/669}

\bibitem[{Weisskopf {et~al.}(1976)Weisskopf, Cohen, Kestenbaum, Novick, Wolff, \& Landecker}]{weisskopf_x-ray_1976}
Weisskopf, M.~C., Cohen, G.~G., Kestenbaum, H.~L., {et~al.} 1976, NASA Special Publication, 389, 81.
\newblock \url{https://ui.adsabs.harvard.edu/abs/1976NASSP.389...81W}

\bibitem[{Weisskopf {et~al.}(2022)Weisskopf, Soffitta, Baldini, Ramsey, O’Dell, Romani, Matt, Deininger, Baumgartner, Bellazzini, Costa, Kolodziejczak, Latronico, Marshall, Muleri, Bongiorno, Tennant, Bucciantini, Dovciak, Marin, Marscher, Poutanen, Slane, Turolla, Kalinowski, Di~Marco, Fabiani, Minuti, La~Monaca, Pinchera, Rankin, Sgro’, Trois, Xie, Alexander, Allen, Amici, Andersen, Antonelli, Antoniak, Attinà, Barbanera, Bachetti, Baggett, Bladt, Brez, Bonino, Boree, Borotto, Breeding, Brienza, Bygott, Caporale, Cardelli, Carpentiero, Castellano, Castronuovo, Cavalli, Cavazzuti, Ceccanti, Centrone, Citraro, D’Amico, D’Alba, Di~Gesu, Del~Monte, Dietz, Di~Lalla, Persio, Dolan, Donnarumma, Evangelista, Ferrant, Ferrazzoli, Ferrie, Footdale, Forsyth, Foster, Garelick, Gunji, Gurnee, Head, Hibbard, Johnson, Kelly, Kilaru, Lefevre, Roy, Loffredo, Lorenzi, Lucchesi, Maddox, Magazzu, Maldera, Manfreda, Mangraviti, Marengo, Marrocchesi, Massaro, Mauger, McCracken, McEachen, Mize, Mereu, Mitchell,
  Mitsuishi, Morbidini, Mosti, Nasimi, Negri, Negro, Nguyen, Nitschke, Nuti, Onizuka, Oppedisano, Orsini, Osborne, Pacheco, Paggi, Painter, Pavelitz, Pentz, Piazzolla, Perri, Pesce-Rollins, Peterson, Pilia, Profeti, Puccetti, Ranganathan, Ratheesh, Reedy, Root, Rubini, Ruswick, Sanchez, Sarra, Santoli, Scalise, Sciortino, Schroeder, Seek, Sosdian, Spandre, Speegle, Tamagawa, Tardiola, Tobia, Thomas, Valerie, Vimercati, Walden, Weddendorf, Wedmore, Welch, Zanetti, \& Zanetti}]{weisskopf_imaging_2022}
Weisskopf, M.~C., Soffitta, P., Baldini, L., {et~al.} 2022, Journal of Astronomical Telescopes, Instruments, and Systems, 8, \dodoi{10.1117/1.JATIS.8.2.026002}

\bibitem[{Wilms {et~al.}(2000)Wilms, Allen, \& McCray}]{wilms_absorption_2000}
Wilms, J., Allen, A., \& McCray, R. 2000, The Astrophysical Journal, 542, 914, \dodoi{10.1086/317016}

\bibitem[{Zane \& Turolla(2006)}]{zane_unveiling_2006}
Zane, S., \& Turolla, R. 2006, Monthly Notices of the Royal Astronomical Society, 366, 727, \dodoi{10.1111/j.1365-2966.2005.09784.x}

\bibitem[{Zane {et~al.}(2001)Zane, Turolla, Stella, \& Treves}]{zane_proton_2001}
Zane, S., Turolla, R., Stella, L., \& Treves, A. 2001, The Astrophysical Journal, 560, 384, \dodoi{10.1086/322360}

\bibitem[{Zane {et~al.}(2023)Zane, Taverna, González–Caniulef, Muleri, Turolla, Heyl, Uchiyama, Ng, Tamagawa, Caiazzo, Lalla, Marshall, Bachetti, Monaca, Gau, Marco, Baldini, Negro, Omodei, Rankin, Matt, Pavlov, Kitaguchi, Krawczynski, Kislat, Kelly, Agudo, Antonelli, Baumgartner, Bellazzini, Bianchi, Bongiorno, Bonino, Brez, Bucciantini, Capitanio, Castellano, Cavazzuti, Chen, Ciprini, Costa, Rosa, Monte, Gesu, Donnarumma, Doroshenko, Dovčiak, Ehlert, Enoto, Evangelista, Fabiani, Ferrazzoli, Garcia, Gunji, Hayashida, Iwakiri, Jorstad, Kaaret, Karas, Kolodziejczak, Latronico, Liodakis, Maldera, Manfreda, Marin, Marinucci, Marscher, Massaro, Mitsuishi, Mizuno, Ng, O’Dell, Oppedisano, Papitto, Peirson, Perri, Pesce-Rollins, Petrucci, Pilia, Possenti, Poutanen, Puccetti, Ramsey, Ratheesh, Roberts, Romani, Sgró, Slane, Soffitta, Spandre, Swartz, Tavecchio, Tawara, Tennant, Thomas, Tombesi, Trois, Tsygankov, Vink, Weisskopf, Wu, \& Xie}]{zane_strong_2023}
Zane, S., Taverna, R., González–Caniulef, D., {et~al.} 2023, The Astrophysical Journal Letters, 944, L27, \dodoi{10.3847/2041-8213/acb703}

\bibitem[{Zhang {et~al.}(2019)Zhang, Santangelo, Feroci, Xu, Lu, Chen, Feng, Zhang, Brandt, Hernanz, Baldini, Bozzo, Campana, De~Rosa, Dong, Evangelista, Karas, Meidinger, Meuris, Nandra, Pan, Pareschi, Orleanski, Huang, Schanne, Sironi, Spiga, Svoboda, Tagliaferri, Tenzer, Vacchi, Zane, Walton, Wang, Winter, Wu, In’ T~Zand, Ahangarianabhari, Ambrosi, Ambrosino, Barbera, Basso, Bayer, Bellazzini, Bellutti, Bertucci, Bertuccio, Borghi, Cao, Cadoux, Campana, Ceraudo, Chen, Chen, Chevenez, Civitani, Cui, Cui, Dauser, Del~Monte, Di~Cosimo, Diebold, Doroshenko, Dovciak, Du, Ducci, Fan, Favre, Fuschino, Gálvez, Gao, Ge, Gevin, Grassi, Gu, Gu, Han, Hong, Hu, Ji, Jia, Jiang, Kennedy, Kreykenbohm, Kuvvetli, Labanti, Latronico, Li, Li, Li, Li, Li, Limousin, Liu, Liu, Lu, Luo, Macera, Malcovati, Martindale, Michalska, Meng, Minuti, Morbidini, Muleri, Paltani, Perinati, Picciotto, Piemonte, Qu, Rachevski, Rashevskaya, Rodriguez, Schanz, Shen, Sheng, Song, Song, Sgro, Sun, Tan, Uttley, Wang, Wang, Wang, Wang, Wang,
  Wang, Watts, Wen, Wilms, Xiong, Yang, Yang, Yang, Yu, Zhang, Zampa, Zampa, Zdziarski, Zhang, Zhang, Zhang, Zhang, Zhang, Zhang, Zhang, Zhang, Zhao, Zheng, Zhou, Zorzi, \& Zwart}]{zhang_enhanced_2019}
Zhang, S., Santangelo, A., Feroci, M., {et~al.} 2019, Science China Physics, Mechanics \& Astronomy, 62, 29502, \dodoi{10.1007/s11433-018-9309-2}

\end{thebibliography}
\bibliographystyle{aasjournal}



\end{document}